\documentclass[preprint]{elsarticle}
\pdfoutput=1

\usepackage{hyperref}
\usepackage{amsmath}
\usepackage{amsfonts}
\usepackage{mathtools}
\usepackage{amssymb}
\usepackage{gensymb}
\usepackage{graphicx}
\usepackage{lmodern}
\usepackage{kpfonts}
\usepackage{algorithm}
\usepackage{algorithmic}
\usepackage{subfigure}
\usepackage{psfrag}
\usepackage[width=0.8\textwidth]{caption}
\usepackage{multirow}
\usepackage{colortbl}

\usepackage{color}

\makeatletter
\newcommand{\mathleft}{\@fleqntrue\@mathmargin0pt}
\newcommand{\mathcenter}{\@fleqnfalse}
\makeatother

\DeclareGraphicsExtensions{.eps,.pdf,.png,.jpg}
\graphicspath{{./Images/}}

\journal{Journal of Computational Physics}

\newcommand{\bx}{\mathbf{x}}

\newcommand{\bJ}{\mathbf{J}}
\newcommand{\bD}{\mathbf{D}}
\newcommand{\bC}{\mathbf{C}}

\newcommand{\bI}{\mathbf{I}}
\newcommand{\bR}{\mathbf{R}}
\newcommand{\bu}{\mathbf{u}}

\newcommand{\bq}{\mathbf{q}}
\newcommand{\bg}{\mathbf{g}}

\newcommand{\bF}{\mathbf{F}}

\newcommand{\bM}{\mathbf{M}}

\newcommand{\bomega}{\boldsymbol{\omega}}
\newcommand{\balpha}{\boldsymbol{\alpha}}

\newcommand{\btau}{\boldsymbol{\tau}}

\newcommand{\beq}[1]{\begin{equation}\label{Chatelain::eq:#1}}
\newcommand{\eeq}{\end{equation}}

\newcommand{\LCFL}{\mathrm{LCFL}}

\RequirePackage[normalem]{ulem} %DIF PREAMBLE
\RequirePackage{color}\definecolor{RED}{rgb}{1,0,0}\definecolor{BLUE}{rgb}{0,0,1} %DIF PREAMBLE
\newcommand{\eq}[1]{Eq.~\eqref{#1}}
\newcommand{\eqs}[1]{Eqs.~\eqref{#1}}
\newcommand{\fig}{Fig.~\ref}

\newcommand{\sect}{Section~\ref}
\newcommand{\sects}{Sections~\ref}

\def\be{\begin{equation}}
\def\ee{\end{equation}}
\def\bea{\begin{eqnarray}}
\def\eea{\end{eqnarray}}
\def\be{\begin{equation}}
\def\ee{\end{equation}}
\def\bea{\begin{eqnarray}}
\def\eea{\end{eqnarray}}

\usepackage{ulem}
\definecolor{darkGreen}{RGB}{0,160,0}
\definecolor{darkBlue}{RGB}{51,51,204}
\definecolor{darkRed}{RGB}{160,0,0}

\begin{document}

\begin{frontmatter}

\title{Simulations of propelling and energy harvesting articulated bodies via vortex particle-mesh methods}
%\title{Coupling a vortex particle-mesh method to a multi-body system solver for the simulation of articulated swimmers}%40 characters

%%% Group authors per affiliation:
%\author{Caroline Bernier\fnref{myfootnote}}

\author[mymainaddress]{Caroline Bernier\corref{mycorrespondingauthor}}
\cortext[mycorrespondingauthor]{Corresponding author.}
\ead{caroline.bernier@uclouvain.be}

\author[mysecondaryaddress,mytertiaryaddress]{Mattia Gazzola}
\author[mymainaddress]{Renaud Ronsse}
\author[mymainaddress]{Philippe Chatelain}

\address[mymainaddress]{Institute of Mechanics, Materials and Civil Engineering (iMMC), Universit\'e catholique de Louvain (UCLouvain), 1348 Louvain-la-Neuve, Belgium}
\address[mysecondaryaddress]{
Department of Mechanical Science and Engineering (MechSE), University of Illinois at Urbana-Champaign (UIUC), Urbana, IL 61801, USA}
\address[mytertiaryaddress]{
National Center for Supercomputing Applications (NCSA), UIUC, Urbana, IL 61801, USA}

%!TEX root = JCP_bernier.tex
 \begin{abstract}
The emergence and understanding of new design paradigms that exploit flow induced mechanical instabilities for propulsion or energy harvesting demands robust and accurate flow structure interaction numerical models. 
In this context, we develop a novel two dimensional algorithm that combines a Vortex Particle-Mesh (VPM) method and a Multi-Body System (MBS) solver for the simulation of passive and actuated structures in fluids.
The hydrodynamic forces and torques are recovered through an innovative approach which crucially complements and extends the projection and penalization approach of Coquerelle et al.~\cite{Coquerelle:2008} and Gazzola et al. \cite{Gazzola:2011}.  
The resulting method avoids time consuming computation of the stresses at the wall to recover the force distribution on the surface of complex deforming shapes. This feature distinguishes the proposed approach from other VPM formulations.
The methodology was verified against a number of benchmark results ranging from the sedimentation of a 2D cylinder to a passive three segmented structure in the wake of a cylinder.
%Principaux résultats
We then showcase the capabilities of this method through the study of an energy harvesting structure where the stocking process is modeled by the use of damping elements.

 \end{abstract}


\end{frontmatter}

%\linenumbers

%!TEX root = JCP_bernier.tex
 \section{Introduction}
%Context: swimming
The thorough comprehension of biological locomotion in fluids promises enhanced performances in a range of engineering applications from underwater vehicles to energy harvesting devices. Swimming organisms at large have refined their senses, morphologies, gaits and mechanical properties to effectively propel and maneuver in a variety of unsteady flow conditions. The interplay between sensing and musculoskeletal architectures only started to be uncovered by biologists and engineers alike \cite{Dabiri:2017}.

% Context: simulation
This interest has spurred a number of research efforts in recent years. %, either experimental, theoretical, or numerical. 
From an experimental perspective, biological observations have been complemented by an increasing number of robotic platforms to test biological hypotesis, designs, and actuation schemes~\cite{Ijspeert:2007,Ijspeert:2014}. At the intersection of these disciplines, one should also mention the very recent foray into the development of hybrid robotic-biological systems with the development of a tissue-engineered ray~\cite{Park:2016}. 
From a theoretical standpoint, the interest in swimming dynamics dates back more than half a century. Theories that consider potential flows include the famous Slender Body Theory (SBT) of Lighthill~\cite{Lighthill:1960}, its extension to large amplitude deformation (Large Amplitude Elongated Body Theory, LAEBT)~\cite{Lighthill:1971}, and recent works on inviscid locomotion by Kanso et al.~\cite{Kanso:2005}. These approaches entail low computational costs, which allow their implementation into model-predictive control schemes in robotics, see e.g. Boyer et al.~\cite{Boyer:2010} and Porez et al.~\cite{Porez:2014}.

The shift to a realistic viscous setting and high fidelity entails higher computational costs and, as a consequence, more complex and efficient models. 
A broad spectrum of direct CFD solvers have been applied to swimming problems: finite differences~\cite{Carling:1998}, Lagrangian multipliers~\cite{Shirgaonkar:2009}, viscous vortex methods~\cite{Eldredge:2006}, block lower-upper symmetric Gauss-Seidel~\cite{Zhang:2008}, Singular-Value Decomposition based Generalized Finite Difference (SVD-GFD)~\cite{Yeo:2010}, finite volumes~\cite{Kern:2006} and Smoothed Particle Hydrodynamic~\cite{Hieber:2008,Kajtar:2008}. Let us mention, the advantages of vortex methods for the application at hand: the compactness of the working variable (vorticity) and the straightforward treatment of unbounded problems.
All of these methods need to deal with two specific challenges (i) handle complex and deforming geometries and (ii) couple the flow problem with the swimmer dynamics in a fluid-structure interaction (FSI) problem.

Two classes of methods can be identified for the first problem. One in which the flow domain accounts for the swimming body in an explicit manner; this has been widely used with unstructured grids that deform and track the boundaries of the swimmer (Arbitrary Lagrangian Eulerian methods--ALE~\cite{Liu:1999}). The second class gathers techniques where the boundaries are represented implicitly and have to be accounted for through either an additional force term in the momentum equations or a modification of the numerical stencil; these include Lagrangian Multipliers \cite{Glowinski:2001}, immersed interface methods \cite{Lee:2003}, immersed boundary methods \cite{Saif-Ullah-Khalid:2016,Park:2016a} and Brinkman penalization methods \cite{Engels:2015,Ghaffari:2015}.
As this work centers around vortex methods, we mention representative works for these two families: the explicit representation of geometries by Eldredge \cite{Eldredge:2008a} and the penalization techniques of~\cite{Coquerelle:2008,Hejlesen:2015} or their more computationally efficient variants in~\cite{Gillis:2017}.

This first classification actually also influences the choice of the coupling technique for the second challenge: the FSI problem. Deforming grids favor the computation and integration of the stresses at the wall, which then allows to perform the fluid-structure coupling either in a weak or strong formulation \cite{Kern:2006,Eldredge:2008a}. The implicit geometry representation of immersed boundary and penalization techniques will be more predisposed to a coupling based on a computation of the net momentum exchange between the fluid and the solid, which can be performed through a projection step as in~\cite{Patankar:2005} and also in~\cite{Coquerelle:2008,Gazzola:2011,Wang:2015} in the context of vortex methods. We note that the contribution in \cite{Wang:2015} handles the interaction between the fluid and the articulated structures through a strong coupling, which accurately captures added mass effects and linkage constraints at the cost of a few iterations per time step. We also mention the related effort of Cottet and Maitre \cite{Cottet:2008} who compute the interaction between a flow and elastic membranes through the elastic energy, itself based on the level-set discretization of the membrane.
 
%Limitations
Almost all these efforts have considered the swimming kinematics to be known a priori, and thus have applied the FSI coupling to the swimmer as a whole, i.e. for the recovery of its rigid motion components. Very few numerical works have included the resolution of the swimmer internal dynamics and more specifically, the actuation needed to execute the desired swimming kinematics, i.e. the constraint efforts to enforce a desired kinematics. These exceptions concern the simplified analytical tools developed for the real-time control of robotic platforms in~\cite{Boyer:2010,Porez:2014}, which cannot account for possible perturbations in the flow such as turbulence, large scale structures shed by an obstacle, or the wakes of other swimmers. This context contributes to the rationale behind the present extension of vortex methods as those already combine computational efficiency, even past complex deforming geometries and the ability to handle inflow perturbations.

The accurate numerical simulation of propulsive actuation strategies in complex flow scenarios constitutes a necessary capability towards uncovering the interplay between flow, sensing and gaits, towards the rational design of engineering applications.
The present work precisely aims at addressing this challenge. 
To that end, we develop a computational framework that unifies the treatments of the flow and of the swimmer. Specifically, we extend the methodology introduced by Coquerelle et al.~\cite{Coquerelle:2008} and further developed by Gazzola et al.~\cite{Gazzola:2011} to allow the kinematics of the actuated device to be solved along with the fluid dynamics. The proposed approach retains the penalization/projection steps of the original works but deploys them in a more general fashion, which allows the extraction of hydrodynamic efforts. These efforts can then be seamlessly integrated into a Multi-Body System (MBS) solver. For complicated structures, kinematic constraints (e.g. rotational joints, linear rails, \textit{etc}) can be enforced exactly by the MBS solver. This waives the use of a stiff forcing term in the flow equations and actually relaxes the stability and accuracy constraints on the time integration within the flow solver.

A central component of the proposed scheme is the computation of the hydrodynamic forces and moments on the individual device components based on information from the penalization/projection approach.
The method thus avoids the computation of local wall stresses and their subsequent integration. Incidentally, we note that the recovery of wall stresses in vortex methods has always been challenging. It indeed entails the evaluation of vorticity at the wall, i.e. in a region of high gradients, and the solution of a specific Poisson equation for the pressure at the wall. Verma et al.~\cite{Verma:2017} used the same FSI-enabled VPM method of~\cite{Coquerelle:2008,Gazzola:2011} as in the present work, together with such a local wall stress computation, in order to calculate the power required for locomotion. The present approach can thus be seen to save one Poisson solution compared to Verma et al.~\cite{Verma:2017} but at the cost of recovering integrated forces only.

%Text presentation
This paper is structured as follows. We recall the Vortex Particle-Mesh (VPM) method for prescribed deformations with an FSI coupling based on projection and penalization techniques~\cite{Gazzola:2011}, and then develop the coupled algorithm in \sect{sec:method}. In \sect{sec:verif}, a verification of numerical accuracy is performed on our method by reproducing several benchmarks~\cite{Gazzola:2011,Eldredge:2008a,Shiels:2001,Eldredge:2008}: sedimentation of a 2D cylinder, flow past an elastically mounted cylinder, free swimming of an articulated fish, and its passive locomotion in the wake of a cylinder. In \sect{sec:harvest}, we apply the present framework to an eel-like energy harvester lying in the wake of a cylinder. We close this paper in \sect{sec:concl} with our conclusions and perspectives.
%!TEX root = JCP_bernier.tex
 \section{Methodology \label{sec:method}}  
The present approach builds upon the method developed in \cite{Gazzola:2011}, and crucially extends it. We recall that this method in its original form can only handle the interaction between a flow and an object whose deformations are prescribed; it is then the total linear and angular momenta of the object that are solved for. The proposed extension allows for the treatment of a system of articulated rigid bodies the deformations of which are unknown a priori, and are recovered by solving its internal dynamics.

%The section is structured as follows. We briefly present the Vortex Particle-Mesh (VPM) method for Fluid Structure Interaction (FSI). We then present the new Algorithm where the the explicit forces component are calculated and applied to a Multi-Body System (MBS) solver. We close this section with a short conclusion summarizing the advantages of such a method.

%\subsection{Vortex method for fluid structure interaction}
% MAttia thinks this is too long, should be cut shorter and maybe put a longer version or the details in appendix
%We first briefly present the original algorithm of \cite{Gazzola:2011} in terms of its components~: a vorticity-based flow solver, the enforcement of wall boundary condition and handling of the fluid-structure interaction. Further details about its implementation are provided in \cite{Gazzola:2011}.

\subsection{The Vortex Particle-Mesh method\label{sec:vpm}}
Both the original work and the present extension rely on a Vortex Particle-Mesh (VPM) method. % for the solution of an 
The incompressible flow past the deforming objects is solved using the velocity ($\bu$)-vorticity ($\bomega=\nabla \times \bu$) formulation of the Navier-Stokes equations (i.e. $\nabla \cdot \bu = 0$)

\begin{align}
\label{eq:vorticity}
\frac{D\bomega}{Dt} & = \left(\bomega\cdot\nabla\right) \bu + \nu \nabla^2 \bomega\\
\nabla \cdot \bu & = 0
\end{align}
where $\frac{D}{Dt}\,=\,\frac{\partial}{\partial t}\,+\,\bu \cdot \nabla$ denotes the Lagrangian derivative, $\bu$ is the velocity field, $\bomega$ the vorticity field and  $\nu$ the kinematic viscosity.
The velocity field is recovered from the vorticity by solving the Poisson equation
\begin{equation}
\nabla^2 \bu =  - \nabla \times \bomega\;.
\label{eq:poisson_u}
\end{equation}
The advection of vorticity is handled in a Lagrangian fashion using particles, characterized by a position $\bx_p$, a volume $V_p$ and a vorticity integral $\balpha_p = \int_{V_p} \bomega d\bx$
\begin{eqnarray}
\frac{d\bx_p}{dt} & = & \bu_p\label{eq:partconv}\\
\frac{d\balpha_p}{dt} & = & \left[(\bomega\cdot \nabla)\bu + \nu \nabla^2 \bomega\right]_{V_p} \;,\label{eq:partdstr}
\end{eqnarray}
where we identify the roles of the velocity field in the advection (\eq{eq:partconv}), and of the vortex stretching and diffusion for the evolution of vorticity (\eq{eq:partdstr}). The stretching term is omitted in the remainder of this article because it is identically equal to zero in a two-dimensional setting. The right-hand sides of these equations are efficiently evaluated on a regular mesh \cite{Chatelain:2008}. The diffusion operator uses second order centered finite difference schemes. The Poisson solver operates in Fourier space for unbounded domain. %and it simultaneously allows for unbounded directions and inlet/outlet boundaries~\cite{Chatelain:2010}. %Inlet/outlet faut voir si on refait les tests avec le nouveau code
To this end, information is made available on the mesh, and retrieved from the mesh, by interpolating back and forth between the particles and the grid using high order interpolation schemes. Advantageously, this hybridization does not affect the good numerical accuracy (in terms of diffusion and dispersion errors) and the stability properties of a particle method. The method adapts the time step according to the diffusion Fourier number defined by  $\Delta t \leq \frac{h^2}{2\nu}$ and to a Lagrangian CFL condition (LCFL) for the explicit time integration of advection~\cite{Koumoutsakos:2005}, e.g. $\Delta t < \LCFL/ \|\nabla \bu\|_{\textrm{max}}$; the latter essentially corresponds to preventing particle trajectories from crossing each other.

\subsection{Brinkman penalization\label{sec:brinkman}}
The immersed objects are described by mollified characteristic functions, $\chi_s$ (see \ref{app:brink}). Each of these functions is built upon a level set function that specifies the signed distance to the surface of the body, $d$. This geometric information is carried by a specific deformable grid associated with the object, and the resulting color function is interpolated onto the regular computational mesh (see \cite{Gazzola:2011} for further details).

The no-slip boundary condition enforcement is performed by means of the Brinkman penalization technique (for a detailed proof of convergence see  \cite{Bost:2010}). This approach extends the fluid domain into the solid region and thus considers the fluid and solid as a global continuous domain $ \Upsilon$, combining the solid domain $\Omega$ and the fluid domain $\Sigma$. The approximation of the no-slip boundary condition is achieved by extending the momentum equation with a term that drives the fluid velocity ${\bf u}$ inside the solid region to a prescribed body velocity~${\bf u}_s$. The penalized Navier-Stokes equations then read
\begin{eqnarray}
%& &\frac{\partial{\bf u}}{\partial t} + {\bf u}\cdot(\nabla {\bf u}) = -\frac{1}{\rho}\nabla p + \nu \nabla^2{\bf u}+\lambda\chi_s({\bf u}_s-{\bf u}),\ {\bf x} \in \Upsilon\label{Eqvelo}
& &\frac{D{\bu}}{D t} = -\frac{1}{\rho}\nabla p + \nu \nabla^2{\bf u}+\lambda\chi_s({\bf u}_s-{\bf u}),\ {\bf x} \in \Upsilon\label{Eqvelo}
%& & \nabla \cdot {\bf u} = 0,\  {\bf x} \in \Upsilon
\end{eqnarray}
where $\lambda\gg 1$ is the penalization factor and $\lambda\chi_s({\bf u}_s-{\bf u})$ is the penalization term. The value of $\lambda$ can be fixed arbitrarily, this directly governs the error in the penalized solution: $||{\bf u}-{\bf u}_\lambda||\leq C \lambda^{-1/2}||{\bf u}||$, as shown in \cite{Carbou:2003}.
A larger value enforces the wall boundary condition more strictly, although at the cost of a stiffer term that needs to be integrated in time.

By taking the curl, we obtain the corresponding penalized velocity-vorticity formulation, 
\begin{equation}
\frac{D\omega}{D t}  = (\omega\cdot\nabla){\bf u} + \nu \nabla^2\omega +\frac{1}{\rho^2}\nabla\rho\times\nabla p +\lambda\nabla\times(\chi_s({\bf u}_s-{\bf u}))\label{eq:vorticity_pen}
\end{equation}
where we identify the additional terms (with respect to \eq{eq:vorticity}) related to the baroclinic generation of vorticity and the penalization. The baroclinic generation of vorticity will appear in the case of a fluid with a space-varying density.  In the present work, the density is considered uniform across the fluid and the solid bodies as we leave the handling of any density difference to the MBS solver. This treatment is allowed by the fact that the effects of a density difference can be well-identified; they are twofold and consist of inertial and gravity effects. 
On the one hand, inertial effects are treated by the use of the specific body density in the integration of the body dynamics by the MBS solver, as discussed in Section \ref{sec:MBS}. 
On the other, the effect of a gravity force can be isolated as it generates a constant hydrostatic pressure gradient in the fluid. 
In the case of a density difference, this gradient in turn generates a constantly active baroclinic vorticity generation term at the fluid-body interface, which can be seen to be equal to the buoyancy force. This means that the buoyancy force can be computed exactly as $(\rho_s-\rho_f) V_s \boldsymbol{g}$ and be simply added to the forces that need to be treated by the MBS. It is worth mentioning that such a treatment is also followed in the Immersed Interface Method (IIM) in velocity-pressure formulation of Xu and Wang \cite{Wu:2010}.
% (which appears in \eq{Eqvelo} and in the baroclinic term of \eq{eq:vorticity_pen}), does have such variations as it represents both the fluid and the immersed solid in a continuous fashion. This field is obtained from the fluid and solid densities, $\rho_s$ and $\rho_f$, through the characteristic function 
%\begin{equation}
%\rho = (1-\chi_s)\rho_f + \chi_s \rho_s\ . 
%\end{equation}

\subsection{Projection technique\label{sec:proj}}
The projection technique captures the effect of the fluid on the solid body. First, at each time step, the fluid evolves freely over the whole domain -- i.e. like if no body was there -- according to the current velocity field. The outcome of this first step is an intermediate state, denoted the \emph{star state}. The resulting extended flow field ${\bf u}_*$ violates the rigid motion of the immersed body. The momentum acquired in this intermediate step by the footprint of the body on the domain correctly captures the flux of momentum between the fluid and the body (for detailed proof see \cite{Patankar:2005}). The captured flux of momentum allows to recover the correct body velocity, which is then used in the penalization step to make the flow field consistent again.

This original algorithm does not explicitly provide the efforts applied to the body but rather their time integral over a time step, i.e. the changes in linear and angular momenta. Unlike the approach of Eldredge \cite{Eldredge:2008}, this approach does not treat the added mass effects implicitly. It thus exhibits a lower convergence in the capture of forces produces by the acceleration of the bodies, as investigated in \cite{Gazzola:2011}. Additionally, the body has prescribed kinematics and the solver does not provide any information about the actuation required to achieve that motion.

\subsection{Extension to an articulated system}
We here modify the algorithm presented in \sects{sec:vpm} to \ref{sec:proj} in order to treat an articulated system of solid elements with embedded actuation. This entails the computation of hydrodynamic forces and moments on each element of the structure to solve the dynamics of the degrees of freedom related to the body joints. In this section, we detail and discuss the components of this extension: the description of the articulated structure, the multi-body system (MBS) solver and the computation of the hydrodynamic efforts from the penalization (\sect{sec:brinkman}) and projection (\sect{sec:proj}) schemes.

\subsubsection{Free articulated system}
\begin{figure}[ht]
\center
		\psfrag{2b}[c][][1.2]{$2b$} 
		\psfrag{2a}[c][][1.2]{$2a$}
		\psfrag{c}[c][][1.2]{$c$} 
		\psfrag{y}[c][][1.2]{$y$}
		\psfrag{x}[c][][1.2]{$x$} 
		\psfrag{q}[c][][1.2]{$q$}
		\psfrag{Rotational}[c][][1.2]{Rotational} 
		\psfrag{joint}[c][][1.2]{joints}
		\psfrag{Linear}[c][][1.2]{Linear}
%\psfragfig[width=10cm]{Images/figure1.eps}
\includegraphics[width=10cm]{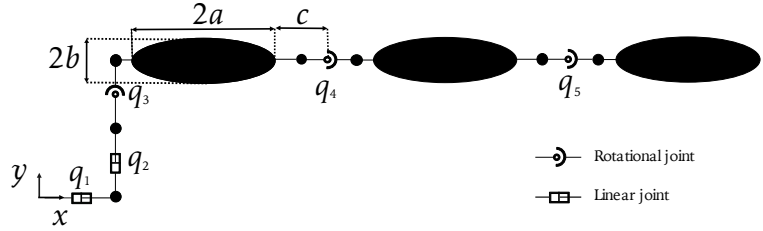}
\caption{Multi-body diagram of three segmented eel-like structure: floating base generalized coordinates, $q_1,\ q_2,\ q_3$, and the joints generalized coordinates,  $q_4,\ q_5$; $2a$ the lenght, $2b$ the width and $c$ the distance from the tip to the hinge joint of each element.\label{fig:MBS1}}
\end{figure}
We focus on a two-dimensional MBS composed of $N$ linked rigid bodies, here elliptical elements, immersed in a fluid (\fig{fig:MBS1}, $N=3$) and connected by means of virtual joints, being here rotational. Each joint entails a degree of freedom that is represented by a generalized coordinate $q_i$, i.e. the angle between two successive elements. 
%betw that reports on the state of the actuator: its position, velocity and acceleration. 
The three first coordinates, $q_1,\ q_2,\ q_3$,  (2 linear and 1 rotational coordinates) describe the floating base, providing the global structure with the three translational and rotational degrees of freedom related to the $ x,\ y$ plane.
The articulated structure is then described by the generalized coordinates: ${\bf q} = [q_1 \ q_2 ... \ q_{n}]^{T}$, with $n=N+2$. The footprint of the structure in space is then described by a set of characteristic functions, $\chi_{s,j}$, each one representing an elliptical element $j$, as in \eq{eq:chi}.

\subsubsection{Multi-body dynamics equations}
\label{sec:MBS}
In this section, we establish the dynamic equations governing for the actuated structure.
% by following the Euler-Lagrange formalism. 
The linear and angular velocities of each body $j$, $\mathbf{v}_{c,j}$ and $\mathbf{\omega}_{c,j}$, are measured with respect to its center of mass and are computed from the generalized coordinates as
\begin{eqnarray}
\mathbf{v}_{c,j} & = & \bJ_{v,j}(\mathbf{q})\,\dot{\mathbf{q}}\\
\mathbf{\omega}_{c,i} & = & \bJ_{\omega,j}(\bq)\,\dot{\mathbf{q}}
\end{eqnarray}
where $\bJ_{v,j}(\bq)$ and $\bJ_{\omega,i}(\bq)$ are the Jacobian matrices associated to the linear and angular velocities, respectively. %\footnote{Here the operator ( $\dot{ }$ ) denotes the time derivative.}. 
Following a classical approach from robotics, we derive the governing equations for the whole system dynamics through the Euler-Lagrange formalism.
The complete dynamical system is then described by the following matrix equation \cite{Spong:2006}
\begin{equation}\label{eqMBS}
\bD({\bf q})\ddot{{\bf q}} +  \bC({\bf q},\dot{{\bf q}})\dot{{\bf q}}= \btau.
\end{equation}
$\bD(\mathbf{q})$ is the inertia matrix:
\begin{equation}
\bD(\mathbf{q}) = \sum^N_{j=1}{\left\lbrace m_i \,\bJ_{v,j}^T \bJ_{v,j} + \bJ_{\omega,j}^T \bR_j(\bq) \bI_j \bR_j^T(\bq) \,\bJ_{\omega,i}\right\rbrace }
\end{equation}
where we dropped the explicit dependencies on $\bq$ of the matrices $\bJ_{v,j}$ and $\bJ_{\omega,j}$ for the sake of clarity. 
The mass of body $j$ and its inertia about a frame fixed to its center of mass are denoted $m_j$ and $\bI_j$, respectively; $R_j({\bf q})$ is the rotation matrix capturing the body orientation with respect to the inertial frame.
The matrix $\bC(\mathbf{q},\dot{\mathbf{q}})$ gathers the Coriolis and centrifugal forces; its elements are given by 
\begin{equation}
C_{lm} = \sum_{i=1}^n \frac{1}{2}\left\lbrace  \frac{\partial D_{lm}}{\partial q_i} + \frac{\partial D_{li}}{\partial q_m} - \frac{\partial D_{im}}{\partial q_l} \right\rbrace \dot{q}_i\;.
\end{equation}
$\btau$ is the effort applied to each joint and is defined as 
\begin{equation}
\btau = \btau_{\textrm{hyd}} + \btau_{\text{act}} \label{eq:torques}
\end{equation}
where $\btau_{\textrm{hyd}}$ is the effort due to the hydrodynamic forces and torques applied to the elements and $\btau_{\textrm{act}}$ is the actuation effort transmitted to the body. For a rotational joint $i$, $\tau_i$ represents the applied torque and for a linear one, it is the force collinear to the joint direction.

%\noindent Construction of each matrix and generalization to $N$ bodies are presented in \ref{Appen1}.

\subsubsection{Fluid-structure interaction efforts}
We now derive the extraction of the hydrodynamic efforts $\btau_{hyd}$ on the structure elements. The original method (\ref{app:former}) solves the FSI problem through the combined use of penalization and projection techniques. As a result, the resulting approach does not explicitly compute the forces exerted on the immersed bodies.

A first departure from the original technique concerns the extension of the flow field inside the solid structure. The fluid properties will be taken as uniform, with viscosity and density values $\nu,\ \rho_f$ everywhere, unlike the original spatially-dependent quantities of \eq{eq:density}. This is motivated by the fact that the whole dynamics of the solid will be handled explicitly by the MBS solver through the hydrodynamic forces $\btau$ in \eq{eqMBS}. The extended domain simply provides a support for the computation of rates of momentum transfer, as shown below.

We recover the hydrodynamic forces and moments from the penalization and projection steps, as shown in \ref{app:hydro}:
\begin{eqnarray}
{\bF}_{\textrm{hyd},j} & = & \frac{d}{dt} \int_{\Omega_{mat,j}}\rho_f \bu\ dV + \int_{\Omega_j}  \rho_f \lambda\chi_s(\bu-\bu_s)\ dV \label{eq:Fhyd}\\
\bM_{\textrm{hyd},j} & = & \frac{d}{dt}  \int_{\Omega_{mat,j}}\rho_f\, \left( \bx\times \bu \right)\ dV + \int_{\Omega_j}{ \bx\times(\rho_f \lambda\chi_s({\bf u}-{\bf u}_s))}\ dV \label{eq:Mhyd}
\end{eqnarray}
The resulting efforts, ${\bF}_{hyd}$ and $\bM_{\textrm{hyd}}$, are each composed of two terms. The first term is associated with the projection step, since the projection scheme effectively performs the time integration of this term. Conversely, the present approach will perform the evaluation of the time derivatives of these volume integrals. The second term is straightforwardly identified as the contribution due to the penalization technique.
Note that the moment $\bM_{\textrm{hyd}}$ is computed with respect to the origin; it still needs to be translated to the body center of mass $\bx_{\textrm{cm}}$ through $\bM_{\textrm{hyd, cm}} = \bM_{\textrm{hyd}} -  \bx_{\textrm{cm}} \times {\bF}_{\textrm{hyd}}$ in order to be used in the MBS solver (\sect{sec:MBS}).

The uniform density extension has additional consequences. Because the baroclinic term in the vorticity evolution equation (\eq{eq:vorticity_pen}), $1/\rho^2 \nabla\rho \times \nabla p$ vanishes, the efforts of \eqs{eq:Fhyd} and \eqref{eq:Mhyd} are missing the hydrostatic component exerted on a body lighter or heavier than the ambient fluid. 
This component will need to be accounted for within the multi-body system solver through the straightforward addition of corresponding forces and moments, i.e. ${\bF}_{\textrm{stat},j} = (\rho_s-\rho_f)\,V_j\,\bg$ and ${\bM}_{\textrm{stat},j} = (\rho_s-\rho_f) V_j \,\bx_{\textrm{cm}} \times \bg$.

\subsubsection{Computational aspects}
%This fluid solver is an hybrid particle-mesh method. It is discretized on a Cartesian computational domain $C$ with an uniform grid spacing $h$. The uniform grid domain is used to evaluate differential operators, to compute velocities and to initialize and re-mesh vortex particles. Spatial derivatives are computed with second order centered finite difference schemes throughout the resolution. % Répétition ?
In the following, we describe the implementation of the above methodology into a full algorithm. We recall that the forces and moments $\bF_{\textrm{hyd},j}$ and $\bM_{\textrm{hyd},j}$ are computed independently for each sub-body.  Each element $j$ of the structure then needs to be characterized by a characteristic function $\chi_{s,j}$. 
For the sake of notation simplicity, the subscript $j$ of the sub-body will be omitted in the remainder of this section. Forces and moments will then have to be understood as the set of forces and moments applied to each system element.

The whole coupled method is summarized in Algorithm \ref{Algo}. The superscript $n$ denotes the evaluation time $t^n$ of fields and variables. We then denote the velocity, vorticity, characteristic function, generalized coordinates, forces, moments and density fields at time $t^n$ as ${\bf u}^n$, $\omega^n$, $\chi_s^n$, ${\bf q}^n$, ${\bf F}^n$, $M^n$ and $\rho^n$, respectively.

The second order differential equation for the MBS (\eq{eqMBS}) is integrated by means of a fourth order Runge-Kutta-Fehlberg method. This method offers error control through an adaptive time step: a local error is estimated from solutions obtained with a fourth and fifth order Runge-Kutta method; the time step is reduced if the error is larger than a threshold value $\xi$. In this work, we used $\xi=10^{-6}$ for all simulations.
The high order integration of our MBS solver is motivated by its stability region: these systems often present little or zero damping and high stiffness. Additionally, the associated computational overhead is affordable given the limited number of variables in the MBS ($\mathcal{O}(10^1 - 10^2)$).
\mathleft
\begin{algorithm}
\caption{Coupled Method\label{Algo}}
\begin{scriptsize}
%\begin{algorithmic}
WHILE {$t^n \leq T_{end}$}\\
 \textit{Step 1 : Velocity update}
\begin{align} 
\nabla^2\psi^n &= -\omega^n \label{eq1}\\
{\bf u}^n &= \nabla\times\psi^n \label{eq2}\\
 ||\nabla{\bf u}^{n}||&\Delta t^{n}\leq LCFL\  \text{and}\ \Delta t^n \leq \frac{h^2}{2\nu} \label{eq3}
\end{align}
 \textit{Step 2 : Projection forces}
%Time integration of the flow field to time $t^n+\Delta t^{n}$, without the MBS solver, nor the penalization (i.e. only step 5 followed by a velocity evaluation); this state is denoted as $\bomega^n_*$ and  $\bf u^n_*$.
\begin{align} 
\frac{\partial \omega^*}{\partial t }&= \nu \nabla^2\omega^* - \nabla\cdot({\bf u}^*\omega^*)\text{ integrated from $t^n$ to $t^{n+1} = t^n+\Delta t^n$} \label{eq:proj}\\
 \nabla^2\psi^{*,n+1} &= -\omega^{*,n+1} \\
{\bf u}^{*,n+1}  &= \nabla\times\psi^{*,n+1}  \\
{\bf p}^{*,n+1} _{proj} &= \int_{\Omega_{mat}}{\rho_f\chi_s^{n}{\bf u}^{*,n+1} }d{\bf x}\ \ \ \ \ \ \ \ \text{with}\ \ \ \ \ \ \ \ \bF^{n+1} _{proj} = \frac{{\bf p}^{*,n+1} _{proj}-{\bf p}^{*,n} _{proj}}{\Delta t^n} \label{eq4}\\
{\bf l}^{*,n+1} _{proj} &= \int_{\Omega_{mat}}{\rho_f\chi_s^{n}\ (\bx\ \times\ {\bf u}^{*,n+1} })\ d{\bf x}+\bx^n_{cm} \times \bF^{n+1} _{proj}\ \ \ \ \ \ \text{with}\ \ \ \ \ \ \bM^{n+1} _{proj} =\frac{{\bf l}^{*,n+1} _{proj}-{\bf l}^{*,n} _{proj}}{\Delta t^n}\label{eq5}
\end{align}
 \textit{Step 3 : Time integration of the MBS}
\begin{align}
 \btau&^{n+1}_{stat} = \mathcal{F}(\bF^{n+1}_{stat},\bM^{n+1}_{stat}) \label{eq:taustat}\\
 \btau&^{n+1}_{hyd} = \mathcal{F}(\bF^{n+1}_{proj}+\bF^{n}_{pen},\bM^{n+1}_{proj}+\bM^{n}_{pen})\label{eq:tauhyd} \\
 \btau&^{n+1}_{act}\text{ provided by a control law} \nonumber \\
D&({\bf q})\ddot{{\bf q}} + C({\bf q},\dot{{\bf q}})\dot{{\bf q}} = \btau^{n+1}_{hydro} + \btau^{n+1}_{act} + \btau^{n+1}_{stat}\\
&\text{Compute }{\bf q}^{n+1}\text{ and }\dot{{\bf q}}^{n+1}\text{ for }t^{n+1}=t^n+\Delta t^n\nonumber\\
\chi&_s^{n+1}=\mathcal{G}({\bf q}^{n+1})\label{G}\\
{\bf u}&_s^{n+1} = \mathcal{H}({\bf q}^{n+1},\dot{{\bf q}}^{n+1}\label{H})
\end{align}
 \textit{Step 4 : Penalization}
\begin{align}
{\bf u}&_\lambda^{n+1} = \frac{{\bf u}^n+ \lambda\Delta t\chi_s^{n+1}{\bf u}_s^{n+1}}{1+\lambda\Delta t \chi_s^{n+1}}\label{eqpen1}\\
\omega&_\lambda = \nabla\times{\bf u}_\lambda^{n+1}\label{eqpen2}\\
\bF&^{n+1}_{pen}=\int_\Omega{\lambda\rho_f\chi_s^{n+1}({\bf u}^{n+1}_\lambda-{\bf u}^{n+1}_s)}d{\bf x}\label{eq:Fpen}\\
\bM&^{n+1}_{pen}=\int_\Omega{\lambda\rho_f\chi_s^{n+1}{\bf x}\times({\bf u}^{n+1}_\lambda-{\bf u}^{n+1}_s)}d{\bf x}+\bx^{n+1}_{cm} \times F^{n+1}_{pen}\label{eq:Mpen}
\end{align}
 \textit{Step 5 : Time integration of the vorticity field}
\begin{align}
\frac{\partial \omega_\lambda}{\partial t }&= \nu \nabla^2\omega_\lambda - \nabla\cdot({\bf u}_\lambda\omega_\lambda) \label{eqdiff}\\
 \omega^{n+1} &= \omega_\lambda^{n+1}\\
t&^{n+1} = t^{n} + \Delta t^{n} \nonumber\\
n& = n+1\nonumber
\end{align}
ENDWHILE

\end{scriptsize}
%\end{algorithmic}
\end{algorithm}
\mathcenter

\paragraph{Step 1: Velocity update}
The velocity is recovered from the Poisson equation (\eqs{eq1} and \eqref{eq2}), solved by an unbounded Fourier solver. The time step is constrained by the diffusion Fourier number and the advection CFL condition. Especially the lagrangian CFL condition \eqref{eq3} is used in order to control the Lagrangian distortion of particles over a time step due to the remeshing of the particles~\cite{Koumoutsakos:2005,Cottet:2000}. % IL FAUT absolument en parler dans les sections avant + en détail ... 

\paragraph{Step 2: Projection forces}
The projection step consists in letting the whole flow (i.e. including the extension inside the body) evolve over a time step $\Delta t$ as if the structure was not there.
The resulting velocity field ${\bf u}_*^n$ corresponds to a gain in linear and angular momentum for each element of the immersed bodies. This predicted momentum is used to compute the flux of momentum between two time steps by performing the time differentiation of \eqs{eq:Fhyd} and \eqref{eq:Mhyd}; these steps correspond to \eqs{eq4} and \eqref{eq5}.

\paragraph{Step 3: Solve the MBS}
The hydrostatic and hydrodynamic torques acting at the joints are obtained through a free body diagram method. They are computed from the applied forces and moments, both hydrostatic (\eq{eq:taustat}) and hydrodynamic (\eq{eq:tauhyd}), through a mapping $\mathcal{F}$. The MBS solver uses these torques and the actuation torques provided by a control law, $\btau_{\textrm{act}}$, to advance the structure to a new configuration ${\bf q^{n+1}}$.

This configuration is then translated into a characteristic function that describes the structure shape at $t^{n+1}$, $\chi_s^{n+1}$. This step is summarized by a mapping function $\mathcal{G}$ in \eq{G}. Similarly, the velocity field of the structure is also found through the function $\mathcal{H}$ in \eq{H}. 
% Expliquer G et H ... est-ce que décrire en annexe ?

\paragraph{Step 4: Penalization}
The no-slip condition is enforced at the interfaces approximately through the penalization of the velocity field. A first order implicit Euler time discretization scheme is used in \eq{eqpen1}.
The penalized vorticity is then recovered from the penalized velocity (\eq{eqpen2}). The forces due to this constraint are computed via the integration of the penalization term over the body surface via an implicit Euler scheme; these penalization forces, \eqs{eq:Fpen} and \eqref{eq:Mpen}, will be used in step 3 of the next time step.

\paragraph{Step 5: Solving the vorticity field}
Step 5 describes the vorticity field update. The vorticity field is diffused and advected through \eq{eqdiff}  with an explicit second order scheme as in the original method~\cite{Gazzola:2011}.
%!TEX root = JCP_bernier.tex
 \section{Verification \label{sec:verif}}
In this section, we verify the methodology derived above by solving several well-documented test cases and compare the performance of our method with that of competing techniques.
These cases include the sedimentation of a 2D cylinder, the flow past an elastically-mounted cylinder, the free swimming of an articulated fish, and the passive propulsion of a fish in a wake.
We use the case of the elastically-mounted cylinder for a convergence study.

\subsection{Sedimentation of a 2D cylinder}
We first test the two-way fluid-solid coupling for a rigid body subject to gravity using the case of a falling 2D cylinder. We compare the results to Gazzola et al.\cite{Gazzola:2011} and Namkoong et al. \cite{Namkoong:2008}.

\subsubsection{Configuration}
A rigid 2D cylinder of diameter $D=0.005\ m$ and with $\rho_s/\rho_f = 1.01$ is released from rest and accelerates due to gravity ($g = 9.81 m/s^2$) until it reaches its asymptotic terminal velocity, corresponding to a $Re = 156$ in water ($\nu = 8\cdot 10^{-7}\,m^2/s, \rho_f = 996\,kg/m^3$). The domain size was set to $[0;8.75\ D]\times[0;70\ D]$ discretized by a $1024 \times 8192$ grid. The following numerical parameters for VPM and the penalization are used: $\LCFL = 10^{-1}$, $\epsilon = 2\sqrt{2}h$, $\lambda = 10^{4}$ with $h$ being the mesh spacing. This resolution of $117$ mesh points per cylinder diameter is similar to the $128$ points used in \cite{Gazzola:2011}.
% $\LCFL = 10^{-1}$, $\epsilon = 2\sqrt{2}h$, $\lambda = 10^{4}$ with $h$ the size of one uniform mesh element taken in the x direction equal to the length of the domain divided by the number of grid point.

\subsubsection{Results}

\begin{figure}
\center
\center
		\psfrag{ylabel}[c][][0.8]{$\mathbf{u}/U_t$} 
		\psfrag{xlabel}[c][][0.8]{$t^*=\frac{tU}{D}$}
\includegraphics[width=0.65\textwidth]{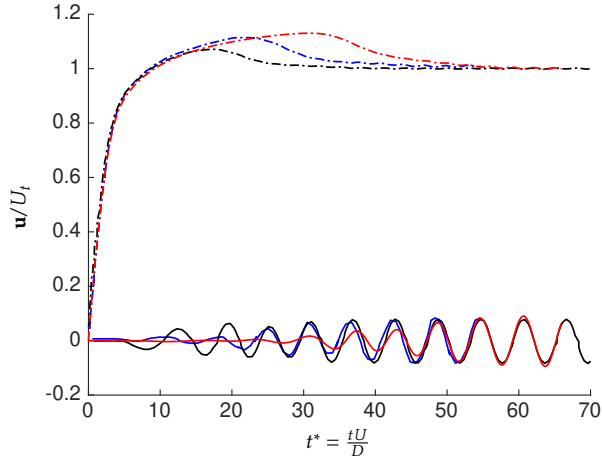}
\caption{Sedimentation of a 2D cylinder: Normalized streamwise velocity $u_y/U_t$ (solid lines) and lateral $u_{lat}/U_t$ (dashed lines); present method (red lines), Gazzola et al. \cite{Gazzola:2011} (blue lines) and Namkoong et al.~\cite{Namkoong:2008} *(black lines).\label{sedi}}
\end{figure}

\fig{sedi} shows the time evolution of the normalized vertical, $u_y/U_t$, and lateral, $u_{lat}/U_t$, velocities obtained by the present method and by the references. $U_t$ represents the reference terminal velocity $U_t = 2.501$ cm/s achieved by \cite{Namkoong:2008}. Similar dynamics are observed. Specifically, the falling velocity overshoots the terminal velocity and then slows down when the shedding starts. The terminal velocity is seen to differ from the reference by less than 1\%. The lateral and angular (not shown) velocities in the steady regime also agree both in amplitude and frequency. The initial transient is triggered by numerical noise when symmetry of the flow is broken. This effect is not controlled and depends on the grid and implementation of the method. This effect explains the delay and overshoot observed for the falling velocity, as well as the transient of the lateral velocity.

\subsection{Flow past an elastically-mounted cylinder \label{flowpastelast}}
\begin{figure}[!ht]
\center
		\psfrag{U}[c][][1.2]{$U$} 
		\psfrag{D}[c][][1.2]{$D$}
		\psfrag{y}[c][][1.2]{$y$} 
		\psfrag{x}[c][][1.2]{$x$}
		\psfrag{k}[c][][1.2]{$k$}
\includegraphics[width=0.45\textwidth]{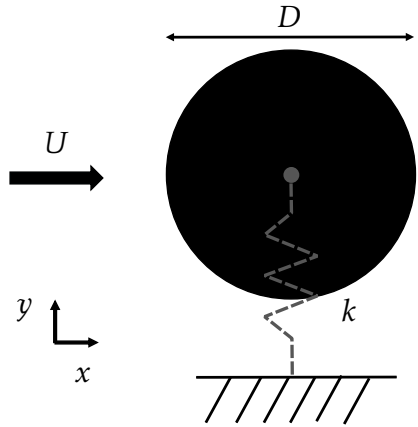}
\caption{Elastically-mounted cylinder in a free stream at $Re= 100$: configuration parameters.\label{cylindergeometry}}
\end{figure}

The vortex shedding generated by the flow past a circular cylinder is a quite common study case in fluid mechanics. Here, we focus on the transverse oscillations induced on an elastically mounted circular cylinder (\fig{cylindergeometry}) due to its vortex shedding, and more specifically on the characterization of these oscillations as the mounting stiffness and the ratio between the fluid and solid densities vary. This validates the handling of the external force, i.e. the elastic spring, on the body by our algorithm as well as the effect of the density difference between the fluid and the body.
%\cite{Shiels:2001}
\subsubsection{Configuration}
We use the problem setup considered in the work of Shiels et al.~\cite{Shiels:2001}, also based on a vortex method but with a panel-based boundary condition enforcement. The Reynolds number is fixed at $Re=\frac{UD}{\nu}=100$, with $D$ the diameter of the cylinder, $U$ the velocity of the induced flow and $\nu$ the kinematic viscosity of the fluid. The cylinder is constrained to move only in the transverse direction, along the $y$-axis. 

The MBS framework presented in \sect{sec:MBS} simplifies into an ordinary differential equation for the cylinder position $y(t)$
\begin{equation}
m\, \ddot{\text{y}}= \bF(t) = F_y(t) - k\,\text{y}(t)\label{motion}
\end{equation}
where the mass coefficient is $m$ and the Coriolis and centrifugal effects vanish because of a single actuated degree of freedom. The total force applied to the system $\bF(t)$ contains the hydrodynamic force $F_y(t)$ and the actuation force $-k \text{y}(t)$, derived from a simple Hooke's law with $k$ the spring stiffness.

We verify the extraction of the hydrodynamic force and the coupling of the flow and MBS solvers through the reproduction of a set of results reported in \cite{Shiels:2001}. It is worth mentioning that this force extraction was also found to agree perfectly with the force measured using the vorticity-based control volume approach of Noca \cite{Noca:1999} (results not shown here for the sake of clarity). The shedding cycle is triggered through a brief starting phase in which the cylinder is rotated with an angular velocity $\Omega = \sin(0.125\pi t^*)$ during the dimensionless time ($t^*=\frac{tU}{D}$) interval $[0;4]$, in a way similar to~\cite{Shiels:2001}. After this period the cylinder is prevented from rotating and is freed in the y direction to undergo the external forces .
% The unsteady wake is induced by the use of a starting phase in which the cylinder is allowed to rotate with an angular velocity $\Omega = \sin(0.25t^*)$ during the dimensionless time interval $t^*=\frac{tU}{D}=8$. After this period the cylinder is prevented to rotate and undergo the hydrodynamic forces in the y direction.

All simulations uses a computational domain size of $[0;21\ D]\times[0;3.5\ D]$ discretized by a $3072 \times 512$ grid. We chose a large domain due to the use of a purely unbounded solver for the Poisson equation and to capture the wake evolution over a long time, thus distance.
The cylinder rest position is located at $(0.63\ D,1.75\ D)$. The following numerical parameters for VPM and the penalization are used: $\LCFL = 10^{-1}$, $\epsilon = h$, $\lambda = 10^{4}$.
\subsubsection{Results}
\begin{figure}[htb]
\center
\subfigure[Tranverse position]{
		\psfrag{ylabel}[c][][0.8][180]{$y/D$} 
		\psfrag{xlabel}[c][][0.8]{$t^*$}
		\includegraphics[width=0.45\textwidth]{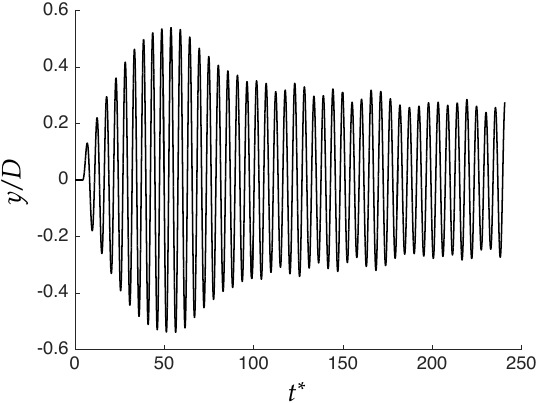}}
\subfigure[Tranverse position zoom]{
		\psfrag{ylabel}[c][][0.8][180]{$y/D$} 
		\psfrag{xlabel}[c][][0.8]{$t^*$}
		\includegraphics[width=0.45\textwidth]{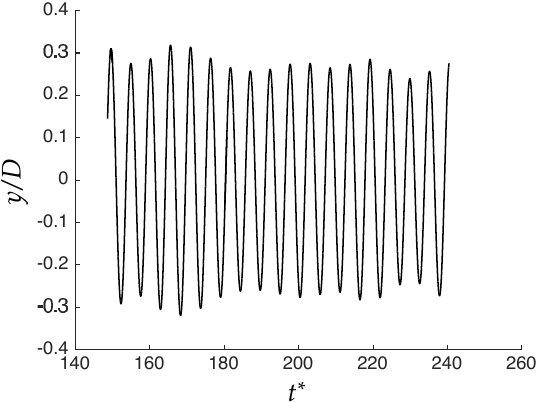}}
\subfigure[Lift coefficient]{
		\psfrag{ylabel}[c][][0.8][180]{$C_l$} 
		\psfrag{xlabel}[c][][0.8]{$t^*$}
		\includegraphics[width=0.45\textwidth]{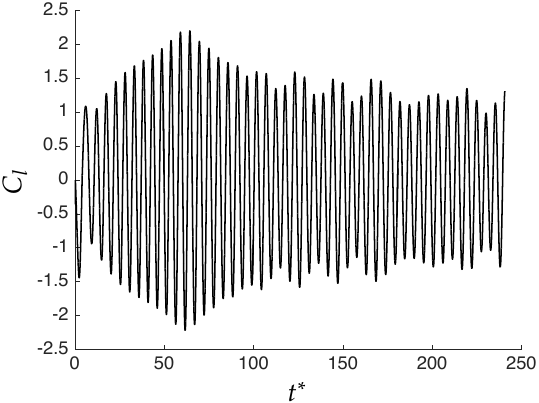}}
\subfigure[Drag coefficient]{
		\psfrag{ylabel}[c][][0.8][180]{$C_d$} 
		\psfrag{xlabel}[c][][0.8]{$t^*$}
		\includegraphics[width=0.45\textwidth]{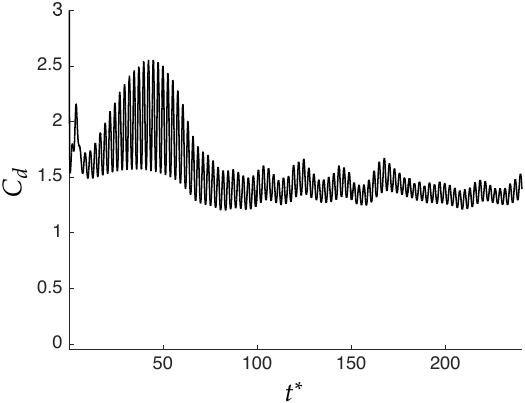}}
\caption{Elastically-mounted cylinder in a free stream at $Re= 100$: case {$m^* = 7.5$} and $k^* = 14.84$, (a,b) trajectory and (c) lift coefficient and (d) drag coefficient, as a function of dimensionless time. \label{N3Results}}
\end{figure}

Figure~\ref{N3Results} shows a representative result in terms of kinematics and dynamics. As expected, the adopted motion converges to a periodic oscillation.
%The drag coefficient oscillates at twice the frequency of the movement. The lift coefficient is then depicted and induced the periodic movement of the cylinder in combination to the spring.

\begin{figure}
\center 
\subfigure[Lift coefficient  $C'_l=\frac{2F'_{y}}{\rho U^2D}$]{
		\psfrag{ylabel}[c][][0.8]{$C'_l$} 
		\psfrag{xlabel}[c][][0.8]{$k_{eff}^*$}
		\includegraphics[width=0.4\textwidth]{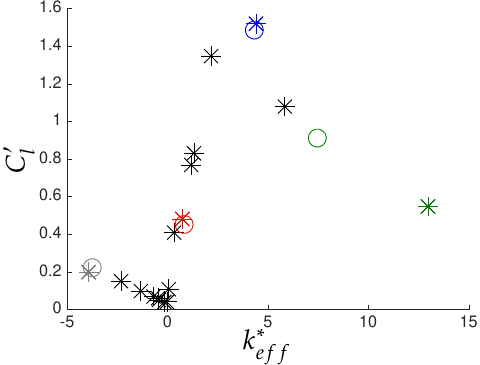}}
\subfigure[Drag coefficient $\overline{C}_d=\frac{2\overline{F}_{x}}{\rho U^2D}$]{
		\psfrag{ylabel}[c][][0.8]{$\overline{C}_d$} 
		\psfrag{xlabel}[c][][0.8]{$k_{eff}^*$}
		\includegraphics[width=0.4\textwidth]{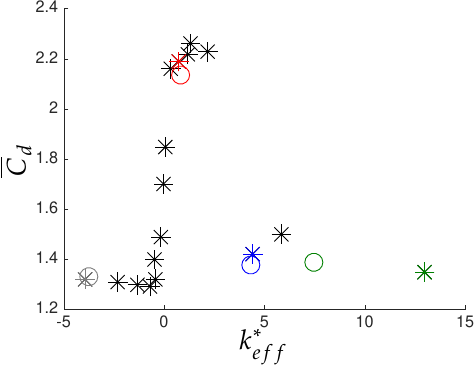}}
\subfigure[Dimensionless amplitude $A^*$]{
		\psfrag{ylabel}[c][][0.8]{$A^*$} 
		\psfrag{xlabel}[c][][0.8]{$k_{eff}^*$}
		\includegraphics[width=0.4\textwidth]{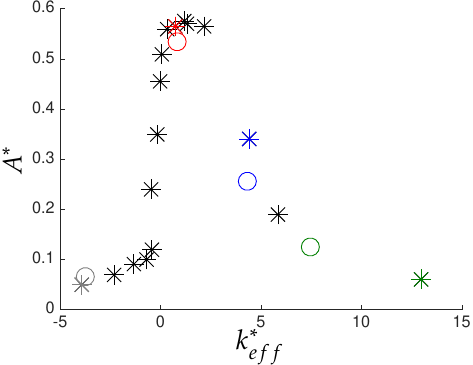}}
\subfigure[Dimensionless frequency $f^*$]{
		\psfrag{ylabel}[c][][0.8]{$f^*$} 
		\psfrag{xlabel}[c][][0.8]{$k_{eff}^*$}
		\includegraphics[width=0.4\textwidth]{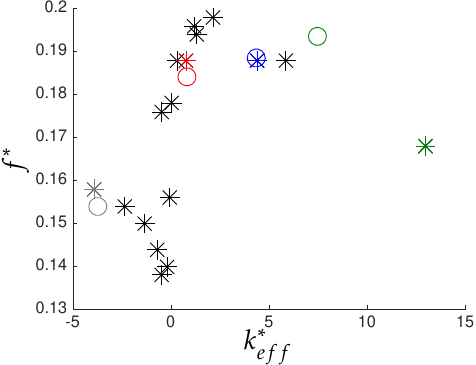}}

\caption{Elastically-mounted cylinder in a free stream at $Re= 100$: dimensionless kinematic and dynamic coefficients: reference data~\cite{Shiels:2001} (star), present results (circles, with matching colors for the same input parameters). \label{recapResults}}
\end{figure}

Figure~\ref{recapResults} and Table~\ref{tab:results} gather several cylinder and stiffness configurations and compare them to the reference~\cite{Shiels:2001}.
Each configuration corresponds to dimensionless numbers $m^* = \frac{\pi\rho_s}{2\rho_f}$ and $k^* = \frac{k}{\frac{1}{2}\rho_fU^2}$ for mass and stiffness, respectively. The resulting cylinder dynamics and kinematics are characterized by the oscillatory part of the lift coefficient $C'_l=\frac{2F'_{y}}{\rho U^2D}$, the mean drag coefficient, $\overline{C}_d=\frac{2\overline{F}_{x}}{\rho U^2D}$, the dimensionless amplitude $A^*=A/D$ and frequency $f^* = f D/U$; these values are estimated from the last fifteen cycles of the simulation results. Following \cite{Shiels:2001}, the effects of inertia and elasticity can be gathered into an effective stiffness  $k_{\text{eff}}^* = (k^*-4\pi^2f^{*2}m^*)$ under the assumption of a purely sinusoidal motion.

The present results are in a satisfactory agreement with the reference. The kinematics and dynamics, in raw form, mostly follow the reference and the global response, as a function of the effective stiffness, reproduces the same highly non-linear response. Our result for the highest mass ratio exhibits a slight departure from the reference. It can be explained by several factors: (i) the computational domain and the simulation time of Shiels are longer ($50$ dimensionless times compared to our $21$); (ii) our simulation (as seen in Fig. \fig{N3Results}) is still settling towards an asymptotic regime: the amplitude $A$ is still decreasing; (iii) the translation of the results to an effective stiffness makes the comparison quite severe as it uses the square of the frequency.

\begin{table}
\center
\begin{tabular}{c|c|c|c|c|c|c|c}
Case &  $k^*$& $k^*_{eff}$ & $m^*$ & $f^*$ & $A^*$ & $C'_l$ & $\overline{C}_d$ \\
\hline \hline Present work &\multirow{2}{*}{0} & -3.74&  \multirow{2}{*}{4} &   0.154 &    0.06 &   0.22 &    1.33\\
 Reference \cite{Shiels:2001} &  & -3.94 &  & 0.158 &   0.05&    0.20&   1.32\\
\hline Present work & \multirow{2}{*}{2.48} & \multirow{2}{*}{0.810} &   \multirow{2}{*}{1.25} & 0.184 &   0.53&    0.45&   2.13\\
 Reference \cite{Shiels:2001} &  &  &  & 0.184 &  0.57&    0.45&   2.16\\
\hline Present work &\multirow{2}{*}{14.84} & 3.81 & \multirow{2}{*}{7.5} & 0.1885 & 0.26  &1.49 &1.38 \\              
 Reference \cite{Shiels:2001} &  & 4.37 &  & 0.188 &   0.34&    1.52&  1.42\\
\hline Present work & \multirow{2}{*}{29.68} & 7.39 &   \multirow{2}{*}{15} &   0.194&    0.12 &    0.91 &   1.39\\ 
 Reference \cite{Shiels:2001} &  & 12.96  &  &    0.168&    0.06&    0.55&   1.35\\
\end{tabular}
\caption{Elastically-mounted cylinder in a free stream at $Re= 100$: dimensionless kinematic and dynamic coefficients.
%: results for the four simulations; spring stiffness $k^*$, effective stiffness $k^*_{eff}$, mass ratio $m^*$, frequency $f^*$, amplitude $A^*$, lift coefficient $C'_l$ and drag coefficient $\overline{C}_d$. 
\label{tab:results}}
\end{table}

\subsubsection{Convergence\label{Sec:error}}
We use the case with $m^* = 1.25$ and $k*=2.48$ to assess the convergence of the method. We compute the solution at $t^* = 0.5$ using a tighter computational domain: $[0;3\ D]\times[0;3\ D]$ as the wake has not developed yet.
%Therefore only one rotation was imposed, namely the rotation of the cylinder that trigger the shedding. 
%Therefore only one degree of freedom was let free to evolve, namely the vertical motion of the cylinder. Simulations were stopped when the cylinder reached its steady motion, which was reached at $t^* = 20$.
We set the ratio $\epsilon/h = 1$ and maintain the Lagrangian CFL at $LCFL = 0.002$. 
This $LCFL$ value was chosen such that the time step was essentially constant across all resolutions, thus maintaining a similar and very low time integration error.
The penalization parameter is kept at $\lambda = 10^4$. We vary the space resolution from $128 \times 128$ to $2048\times 2048$, which we use as our reference solution.

The convergence order was determined by considering the $L^2$ and $L^\infty$ norms of the error $e(\mathbf{x})$ of the vorticity field with respect to the reference solution
\begin{equation}\label{eq:erreur}
e(\mathbf{x}) = ||\omega(\mathbf{x})-\omega_{\text{reference}}(\mathbf{x})||.
\end{equation} 

\begin{figure}
\center
		\psfrag{ylabel}[c][][0.8]{$L^2,L^\infty$} 
		\psfrag{xlabel}[c][][0.8]{Resolution}
\includegraphics[width=0.65\textwidth]{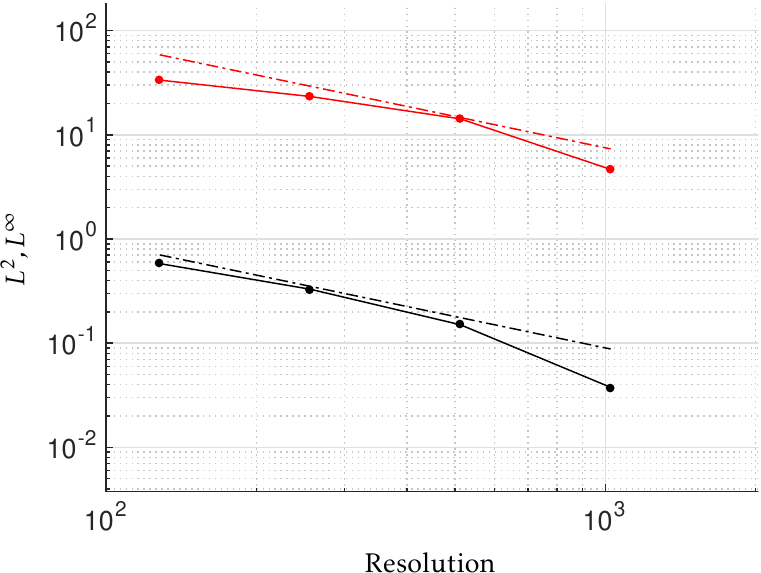}
\caption{Elastically-mounted cylinder in a free stream at $Re= 100$, convergence study: $L^\infty$ (red) and $L^2$ (black) errors; dashed lines indicate first order.\label{error}}
\end{figure}
%Space convergence studies were carried out fixing the model, i.e. by fixing the ratio $\epsilon/h$, to investigate convergence to the actual geometry. For both convergence studies, we set LCFL = 0.001 (to be LCFL-bound at all times), $\lambda = 10^4$ and varied the resolution between $128\times 128$ and $2048\times 2048$. 
\noindent Both these errors (shown in \fig{error}) exhibit first order convergence; this behavior is consistent with the mollification of the immersed object (using a fixed  $\epsilon/h$) in the penalization technique. 

\subsection{Articulated eel-like swimmer}
We verify the handling of a chain of several elements with the case of an eel-like self-propelled swimmer. The contributions of Eldredge~\cite{Eldredge:2008a} and Kanso et al.~\cite{Kanso:2005} who assembled swimmers from articulated ellipses with prescribed kinematics are taken as references. This setup allows to assess the extraction of hydrodynamic forces within a MBS and the application of joint kinematics to a complex structure.

\subsubsection{Configuration}
We consider a swimmer composed of three ellipses with an aspect ratio of $\frac{a}{b} = 10$. The distance that separates a hinge from the ellipse tips $c$ is taken equal to $0.2\,a$ (see \fig{fig:MBS1}). %Because the references use prescribed kinematics and our method is actuation -based, a control layer is added in order to generate the actuation torque signals required for the given kinematics 
We impose the same prescribed kinematic patterns as in the references
\begin{equation}
q_4(t) = -\cos(t-\pi/2)\ \ \text{and}\ \ q_5(t) = -\cos(t).
\end{equation}
The initial velocities are chosen to match the ones of the reference article by taking the same initial values $\dot{q}_{i, \textrm{init}}$.  Our simulations match the \emph{undulation} Reynolds number proposed by Eldredge \cite{Eldredge:2008a}, based on the peak joint angular velocity $\dot{q}_{max}$ and the total length of one ellipse, i.e. $Re =\frac{\dot{q}_{max}(2a)^2}{\nu} = 200$, whereas the results of Kanso et al.~\cite{Kanso:2005} provide a theoretical inviscid reference as they are based on the theory for slender bodies. 
Furthermore, we note that both these references assume massless bodies while the present study considers them as neutrally-buoyant. We still expect the comparison to be relevant as inertial effects are dominated by the added masses of the ellipses along their minor axes ($m_{22} = \rho_f \pi a^2$); the ellipse mass difference ($m=\rho_f \pi a b$ instead of $m = 0$) can then be seen to be a $10\%$ discrepancy.

%While in the present study, the body is neutrally buoyant ($\rho_f = \rho_s$), the references considered massless bodies. Furthermore, Kanso et al.~\cite{Kanso:2005} used the inviscid theory for slender bodies. We match the \emph{undulation} Reynolds number proposed by~\cite{Eldredge:2008a}, based on the peak joint angular velocity $\dot{q}_{max}$ and the total length of one ellipse, i.e. $Re =\frac{\dot{q}_{max}(2a)^2}{\nu}$. 
%The common Reynolds number in this study is $Re = 200$. 
The characteristic length and velocity scales used for the nondimensionalization are respectively $L= 2a$ and U = $\dot{q}_{max}L$.
The simulation uses a computational domain size of $[0;2.04\ L]\times[0;1.53\ L]$ discretized by a $768 \times 576$ grid. 
The swimmer starting position is located at $(0.48\ L,1.06\ L)$. 
The following numerical parameters for VPM and the penalization are used: $\LCFL = 2.0\,10^{-2}$, $\epsilon = 2\,h$, $\lambda = 1.0\,10^{4}$.

\subsubsection{Results}
\begin{figure}
\center
		\psfrag{ylabel}[c][][0.8]{$y/L$} 
		\psfrag{xlabel}[c][][0.8]{$x/L$}
\includegraphics[width=0.65\textwidth]{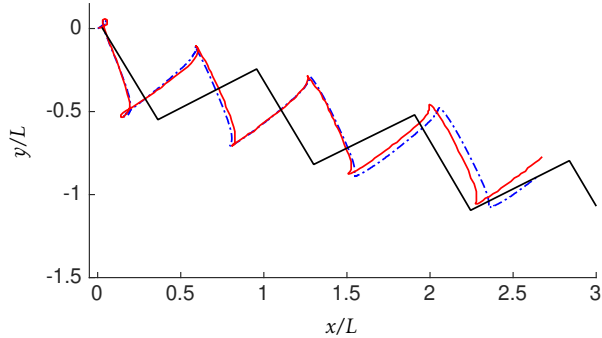}
\caption{Articulated eel-like swimmer:  trajectory of the central body centroid: % present results (solid blue),
 rotated present results (dashed blue), viscous reference~\cite{Eldredge:2008a} (solid red) and inviscid reference~\cite{Kanso:2005} (solid black).\label{Displa}}
\end{figure}

The resulting motion over the course of four undulation periods are compared to \cite{Eldredge:2008a} and \cite{Kanso:2005} in \fig{Displa}. The path of our swimmer was rotated so that its asymptotic direction matches the one of the viscous reference~\cite{Eldredge:2008a}. As a consequence, the steady-state behavior and the initial error can be assessed separately.
This correction corresponds to an angle of $5.14\degree$. This mismatch has two potential causes: a more accurate treatment of added mass effects and a body mass exactly zero in the references. The added mass effects are indeed significant and govern the initial motion of the swimmer and we recall that the original method, on which our algorithm is based, only exhibited first order temporal convergence for added mass effects \cite{Gazzola:2011}.
Concerning the density, numerical tests with a smaller density $\rho_s = 0.5\, \rho_f$(not shown here) indicate that this angular mismatch in the asymptotic trajectory reduces to $3.6^o$, thus hinting at a substantial role of the body density.
The steady motion appears to be in very good agreement with the viscous reference, both regarding the longitudinal velocity and the lateral oscillations, including their shape and the cusp-like features at the direction changes. The inviscid reference produces a faster swimmer with very similar lateral oscillations. The reason is twofold: (i) viscous effects play an important role in the longitudinal dynamics of the present results and the viscous reference~\cite{Eldredge:2008a} and (ii) the lateral dynamics are dominated by potential flow effects, which are correctly predicted by the inviscid reference.
\begin{figure}
\center
\subfigure[Position]{
		\psfrag{ylabel}[c][][0.8]{$x_2,\ y_2,\ \omega_2$} 
		\psfrag{xlabel}[c][][0.8]{$t^*$}
		\includegraphics[width=0.45\textwidth]{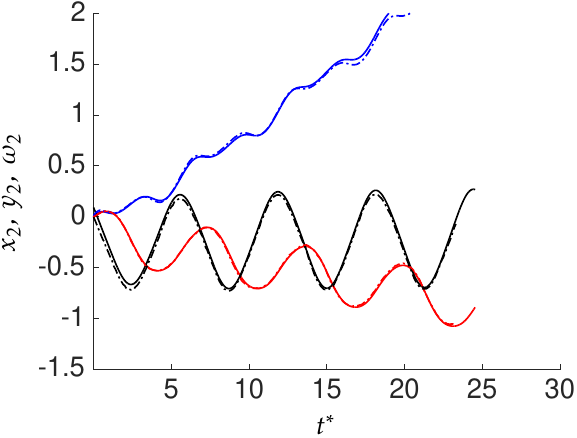}}
\subfigure[Velocity]{
		\psfrag{ylabel}[c][][0.8]{$V_{x,2},\ V_{y,2},\ \Omega_{2}$} 
		\psfrag{xlabel}[c][][0.8]{$t^*$}
		\includegraphics[width=0.45\textwidth]{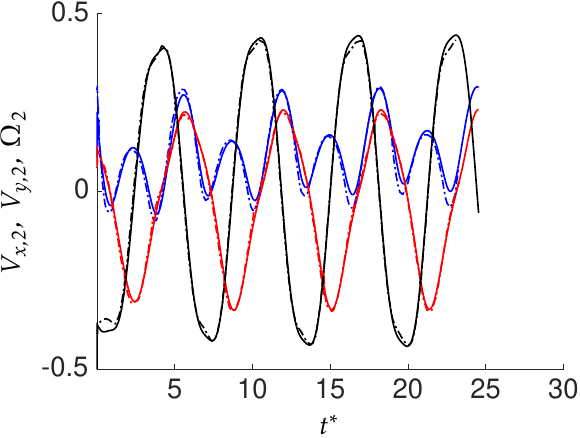}}
\caption{Articulated eel-like swimmer: velocity components of central body; %rotated 
present results (solid), viscous reference~\cite{Eldredge:2008a} (dashed). Position (velocity) plot depicts $x_2$ ($V_{x,2}$): in blue,
$y_2$ ($V_{y,2}$): in black and $\omega_2$ ($\Omega_2$): in red. \label{velo}}
\end{figure}

\begin{figure}[!ht]
\center
		\psfrag{Section1}[c][][0.8]{Section 1} 
		\psfrag{Section2}[c][][0.8]{Section 2} 
		\psfrag{Section3}[c][][0.8]{Section 3} 
		\psfrag{ylabel1}[c][][0.8]{$C_{Fx}$} 
		\psfrag{ylabel2}[c][][0.8]{$C_{Fy}$} 
		\psfrag{ylabel3}[c][][0.8]{$C_{Mz}$} 
		\psfrag{xlabel}[c][][0.8]{$t^*$}
\includegraphics[width=0.9\textwidth]{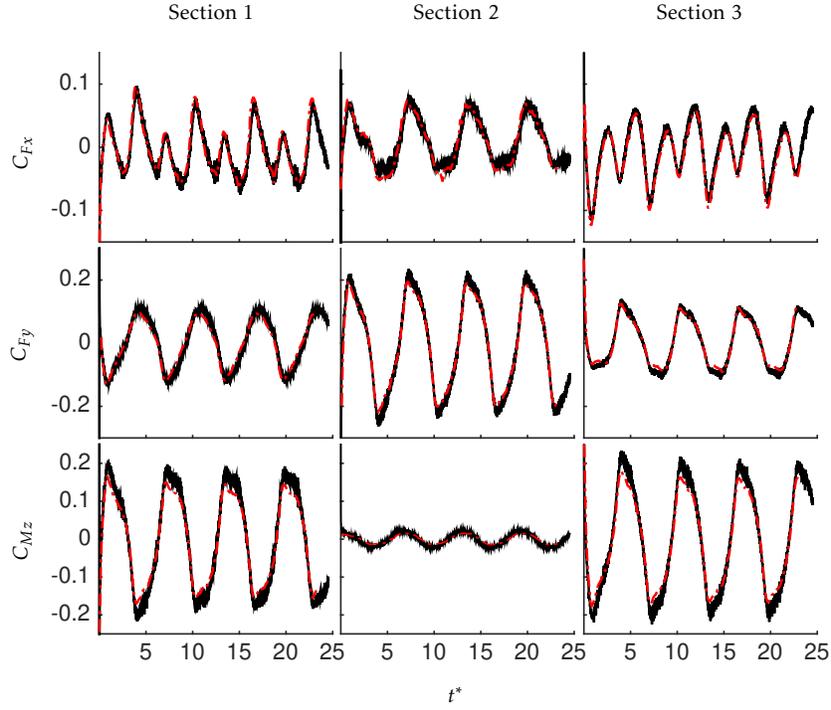}
\caption{Articulated eel-like swimmer: histories of the forces and moments coefficients; %rotated
 present results (solid black), viscous reference~\cite{Eldredge:2008a} (dashed red).\label{forces}}
\end{figure}

\fig{velo} offers further insight into the resulting kinematics of the swimmer and confirms the good agreement with the viscous reference in the steady state regime.
% \pcnote{do we put a zoom into the starting}
%The centroid coordinates, the central-body angle, and their rates of change are plotted in Fig. \ref{velo}. Both simulations are very close to each other and in both simulation the swimmer reaches a periodic steady-state regime.  %The rms velocity of this centroid is 0.238 and its peak speed is 0.373; the centroid travels approximately 0.72 chord-lengths (0.21 body-lengths) per period, at an angle of approximately 15° below the positive x axis.
Forces and moments exerted by the fluid on each constituent body are shown in \fig{forces}. Force histories confirm the excellent agreement with the reference in spite of the different densities for the swimmers (null in the case of~\cite{Eldredge:2008a}). Again, we recall that the present slender geometries cause the added mass effects to overwhelm the inertia effects and dominate the dynamics. Finally, let us mention that the forces are presented in terms of their raw data for the present work, unlike for the reference where some smoothing is applied~\cite{Eldredge:2008a}; the noise in the forces extracted from the penalization and projection steps thus appear quite moderate.

\subsection{Passive propulsion in vortex wakes\label{sec:passive}}
This fourth test case considers an articulated swimmer similar to the one used in the third test case. The swimmer is initially set in a uniform flow past a cylinder as depicted in \fig{Wake} and its joints are free to rotate, i.e. there is no actuation torque.
This setup allows to validate the influence of upstream perturbations on the behavior of a complex system and the application of kinematic constraint between serial body joints.

%This setup allows to assess the extraction of hydrodynamic forces within a MBS and the application of joint constraints.
\subsubsection{Configuration}
\begin{figure}[!ht]
\center
		\psfrag{U}[c][][0.8]{$U$} 
		\psfrag{D}[c][][0.8]{$D$} 
		\psfrag{d}[c][][0.8]{$d$} 
		\psfrag{L}[c][][0.8]{$L$} 
		\psfrag{h}[c][][0.8]{$H$} 
		\psfrag{y}[c][][0.8]{$y$} 
		\psfrag{x}[c][][0.8]{$x$}
\includegraphics[width=0.9\textwidth]{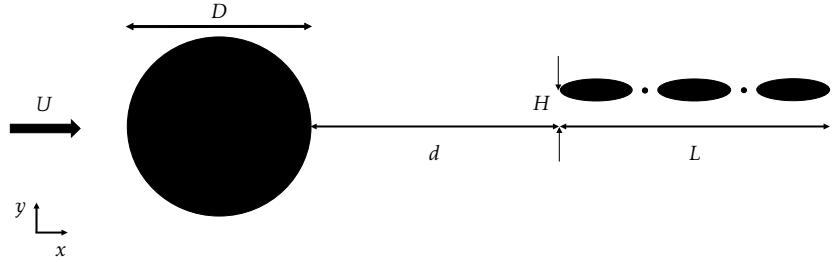}
\caption{Passive propulsion in vortex wakes: configuration setup; cylinder of diameter $D$ and a three-link passive swimmer of length $L$.\label{Wake}}
\end{figure}

We use the problem setup considered in the work of Eldredge~\cite{Eldredge:2008}. A cylinder of diameter $D$ and a three segmented swimmer are immersed in a uniform free stream with a velocity $U$ (see \fig{Wake}). The passive swimmer is neutrally buoyant and placed at a distance $d$ downstream of the cylinder. In order to break the symmetry, the fish is placed above the central line at a distance $H = 0.25\ D$. The swimmer is composed of three identical ellipses, $a =0.3125\ D$,  of aspect ratio $5$, and the distance from tip to hinge \textit{c} is set to $0.1D$. The separation distance $d$ is fixed at $3D$. The fish is constrained to stay behind its initial position in the horizontal direction to prevent it from being swept away due to the incident flow.  The Reynolds number, based on cylinder diameter and free-stream velocity, is taken as $Re= \frac{UD}{\nu}=100$.

The simulation uses a computational domain size of $[0;9.9\ D]\times[0;4.95\ D]$ discretized by a $1536 \times 768$ grid. The cylinder rest position is located at $(1.485\ D,2.475\ D)$. The following numerical parameters for VPM and the penalization are used: $\LCFL = 2.0\,10^{-2}$, $\epsilon = 2h$, $\lambda = 1.0\,10^{4}$.

\subsubsection{Results}
\begin{figure}[!ht]
\center
\subfigure[$tU/D$ = 13.76]{\includegraphics[width=5.5cm]{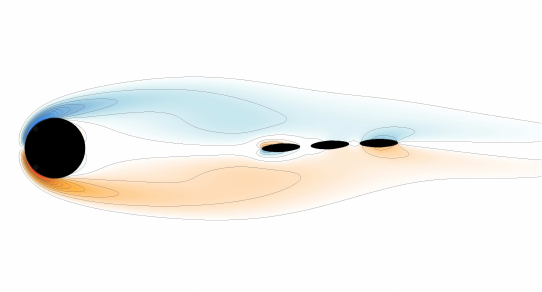}}
\subfigure[$tU/D$ = 24.64]{\includegraphics[width=5.5cm]{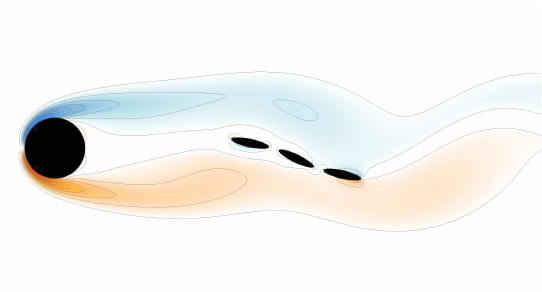}}
\subfigure[$tU/D$ = 26.4]{\includegraphics[width=5.5cm]{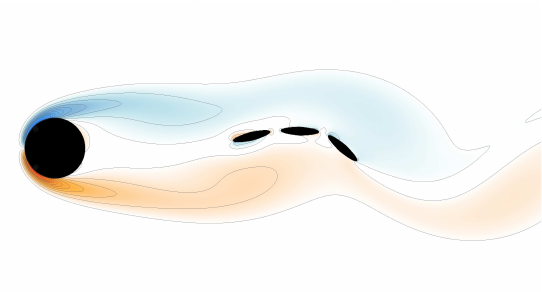}}
\subfigure[$tU/D$ = 28.08]{\includegraphics[width=5.5cm]{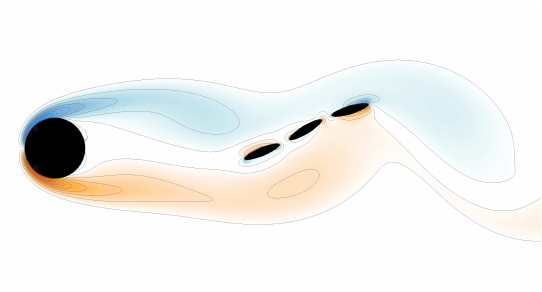}}
\caption{Passive propulsion in vortex wakes: snapshot of the configuration and the vorticity around the free swimmer at various  $t^* = tU/D$. Dimensionless vorticity contours have values from -10 to 10 in 20 uniform increments with negative and positive vorticity in blue and red respectively. \label{vort}}
\end{figure}

Figure~\ref{vort} depicts the vorticity field and the instantaneous configuration of the system for a fish of length $L = 2.275D$. The joints of the swimmer are free to rotate without frictional or elastic resistance. 

\begin{figure}[!ht]
\center
\subfigure[Horizontal position]{
		\psfrag{ylabel}[c][][0.8]{$x/D$} 
		\psfrag{xlabel}[c][][0.8]{$t^*$}
		\includegraphics[width=5.5cm]{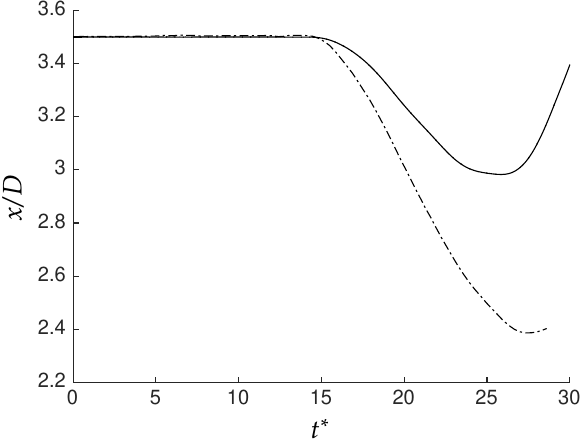}\label{dist}}
\subfigure[First angle of inclination]{
		\psfrag{ylabel}[c][][0.8]{$\theta_1$} 
		\psfrag{xlabel}[c][][0.8]{$t^*$}
		\includegraphics[width=5.5cm]{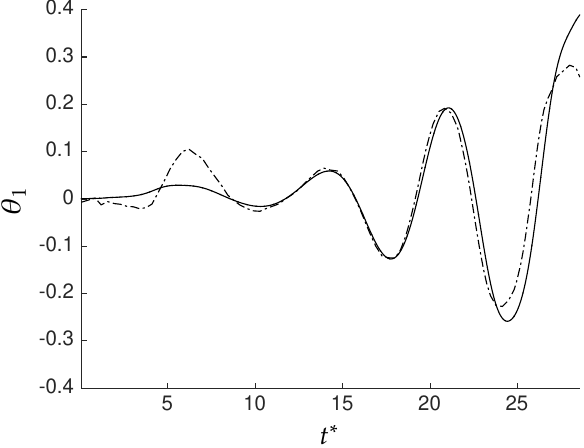}}
\subfigure[Second angle of inclination]{
		\psfrag{ylabel}[c][][0.8]{$\theta_2$} 
		\psfrag{xlabel}[c][][0.8]{$t^*$}
		\includegraphics[width=5.5cm]{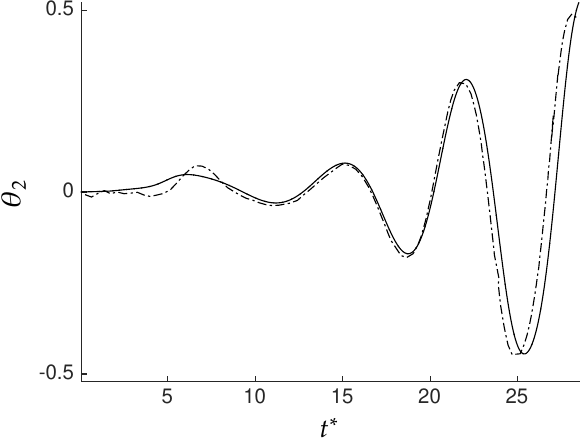}}
\subfigure[Third angle of inclination]{
		\psfrag{ylabel}[c][][0.8]{$\theta_3$} 
		\psfrag{xlabel}[c][][0.8]{$t^*$}
		\includegraphics[width=5.5cm]{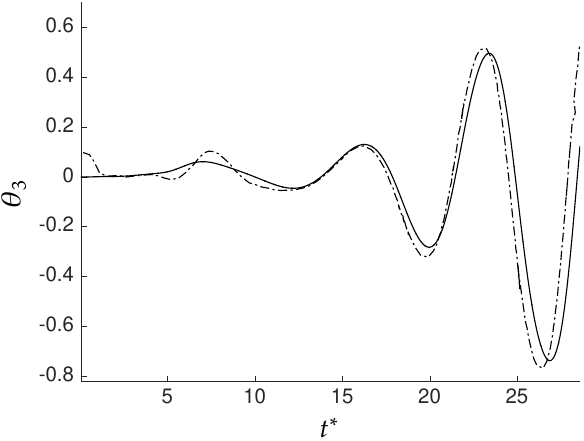}}
\caption{Passive propulsion in vortex wakes: horizontal position of the left-most point of the articulated chain; angle of inclination of each body along the fish with respect to the x axis $\theta_i$; present results (solid), reference~\cite{Eldredge:2008} (dashed).\label{evol}}
\end{figure}

After the vortical wake of the cylinder reaches and surrounds it, the fish moves downward towards the central line. At approximately $t^* = tU/D=13.76$, the fish begins to move towards the cylinder (Fig. \ref{dist}). The reference and the present result begin to move towards the cylinder at the same time, although our swimmer does not move as far upstream. 
This quantitative mismatch is due to the fact that this forward movement is affected by a flow instability which break the symmetry of the wake. The onset of this flow instability is very sensitive to the numerical method used; a sensitivity to the spatial and temporal resolutions was also confirmed by numerical experiments not shown here.
When the fish begins to move towards the cylinder, the symmetry of the wake is broken, and alternating von K\`arm\`an vortex shedding begins. %The fish is stably trapped between the vortices for the entire duration of the simulation.
%\begin{figure}[!ht]
%\center
%		\psfrag{ylabel}[c][][0.8]{$x/D$} 
%		\psfrag{xlabel}[c][][0.8]{$t^*$}
%\includegraphics[width=8cm]{Images/figure8.pdf}
%\caption{Passive propulsion in vortex wakes: horizontal position of the left-most point of the articulated chain; present results (solid) reference~\cite{Eldredge:2008} (dashed).\label{dist}}
%\end{figure}

The induced vortical flow affects the shape of the swimmer and generates an undulation of the articulated body. The histories of the angles of the individual bodies are plotted in Fig. \ref{evol}; they offer a fairer and more meaningful comparison as the swimmer responds to similar flow structures once the vortex shedding has started. The simulations are in a quantitative agreement and one can observe a traveling deformation wave through the phase shift of the consecutive joints. 

%\subsection{Discussion}
%In conclusion, we have shown 3 complementary examples to validate our coupled solver. Test n$\degree $1 enabled us to validate the effect of the actuation on the body via the use of a spring as well as the effect of a density difference between the fluid and the body. Test n$\degree $2 made it possible to validate the hydrodynamic forces transmitted to a complex structure. And finally, the last test allowed to validate a two-way coupling where the body interfered passively to the effect of the fluid.

%!TEX root = JCP_bernier.tex
 \section{Application : energy harvesting\label{sec:harvest}}
 % Petite intro :
In this section, we investigate an energy harvesting application. It is showed that the present FSI technique can be used to simulate complex interactions between an articulated structure and a vortical flow. 
 
% - Choix d'un test où l'on peut extraire de l'énergie de l'environnement
% -Mis en oeuvre par l'utilisation d'amortisseur
 
 \subsection{Configuration}
 This application relies on nearly the same setup as for the validation of Section \ref{sec:passive} (\fig{Wake}). We consider a three-body swimmer immersed in the wake of a cylinder. The novel structure includes passive mechanical elements. As in the previous test case, the Reynolds number, based on the cylinder diameter $D$ and free-stream velocity $U$, is taken to be $Re= \frac{UD}{\nu}=100$.
To restrain the degrees of freedom, we reduce the floating base of the swimming structure by removing the first generalized coordinate $q_1$. The swimmer is fixed at $d=3D$ in the x direction. The other generalized coordinates are set up in parallel with passive elements such as dampers and spring. 

\begin{figure}
 \center
		\psfrag{q}[c][][0.7]{$q$} 
		\psfrag{C}[c][][0.7]{$C$} 
		\psfrag{y}[c][][0.6]{$y$} 
		\psfrag{x}[c][][0.6]{$x$}
		\psfrag{k}[c][][0.7]{$k$}
 \includegraphics[width=11cm]{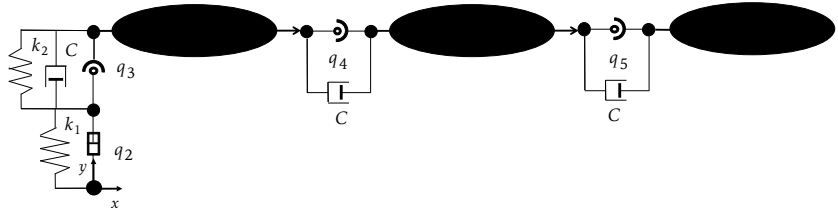}
 \caption{Energy harvesting in vortex wakes: multi-body diagram of 3 bodies eel-like structure with damping. \label{fig:damping}}
 \end{figure}

Figure~\ref{fig:damping} shows the new setup and illustrates the parallel springs $k_1$ and $k_2$ and the dampers $C$. The initial vertical eccentricity of the swimmer with respect to the cylinder $H$ and the initial angular position of the joints are set to zero. 
These damping elements are in charge of carrying out the energy harvesting process. Indeed, damping elements dissipate energy through a reactive torque proportional to the angular velocity of the joint: $\btau_{act}= C\dot{q}_i$, where $C$ is the damping coefficient. 
The parallel dampers added to each joint are described via a coefficient value $C$ common to the three damping elements. A new dimensionless coefficient captures this damping: $C_g = \frac{2\ C}{\rho UD^3}$. 
As in the first verification case, the shedding cycle of the induced flow is triggered via a brief starting phase in which the cylinder is rotated (\sect{flowpastelast}).
 
Here, we investigate the interaction between the harvester and the vortical structures depending on the damping coefficient. We used several damping values: $C_g =\ 25,\ 5,\ 2.5,\ 0.5,\ 0.25,\ 0.05,\ 0.025,\ 0.005$ and $0$. The spring values are described by the following spring coefficients: $k^*_i = \frac{2k_i}{\rho U^2}$ for linear spring and $k^*_i = \frac{2k_i}{\rho U^2 D}$ for angular spring. For the simulations, we choose $k^*_1 = 0.8$ and $k^*_2 = 0.1$. The simulation uses a computational domain size of $[0;14.85\ D]\times[0;4.95\ D]$ discretized by a $1536 \times 768$ grid. The cylinder rest position is located at $(4.45\ D,2.475\ D)$. The following numerical parameters for the VPM method and the penalization are used: $\LCFL = 2.0\,10^{-2}$, $\epsilon = 2h$, $\lambda = 1.0\,10^{4}$.
 %- Reprendre la même configuration que dans le dernier cas de validation tout en y injectant un amortisseur en parrallèle sur chacune des articulations du poisson 
 
\subsection{Results}%envelop for each test ? one ?
In order to describe the results, several dimensionless coefficients are introduced: 
\begin{itemize}
\item the drag coefficients of the cylinder, the harvester and the system as a whole: $C_{D,\textrm{cyl/har/sys}} = \frac{2F_{D,\textrm{cyl/har/sys}}}{\rho U^2D}$ and where $F_{D,\textrm{cyl/har/sys}}$ is the applied streamwise hydrodynamic force; 
\item the lift coefficients of the cylinder, the harvester and the system as a whole: $C_{L,\textrm{cyl/har/sys}} = \frac{2F_{L,\textrm{cyl/har/sys}}}{\rho U^2D}$ where $F_{L,\textrm{cyl/har/sys}}$ is the applied transverse hydrodynamic force; note that the structural force applied on the harvester is recovered from $$\left(F_{D,\textrm{har}}, F_{L,\textrm{har}} \right) =\btau_\textrm{hyd} -\left(\bD({\bf q})\ddot{{\bf q}} +  \bC({\bf q},\dot{{\bf q}})\dot{{\bf q}}\right)\,;$$
\item the mean power coefficient recovered by the damping elements : $\overline{P}^*_{harvest} = \frac{2C\sum_{i=3}^5 \overline{\dot{q}_i}^2}{\rho U^3 D}$;
\item the dimensionless mean power required to tow the cylinder-harvester at a velocity $U$: $\overline{P}^*_{F} = \frac{2(\overline{F}_{D,\textrm{cyl}}+\overline{F}_{D,\textrm{har}})}{\rho U^2 D} =  \frac{2(\overline{F}_{D,\textrm{sys}})}{\rho U^2 D}$; note that this is identical to $C_{D,\textrm{sys}}$;
\item the harvesting efficiency: $\eta_h = \frac{\overline{P}^*_{harvest}}{\overline{P}^*_F}$.
\end{itemize}

\subsubsection{Harvesting performances}
\begin{figure}[bth] 
\center
\psfrag{x}[r][][1.2]{$C_{g}$} 
\psfrag{y1}[r][][1.2]{$\overline{P}^*$}
\psfrag{y2}[r][][1.2]{$\eta_h$}
\psfrag{Q3}[r][][0.6]{$q_3$}
\psfrag{Q4}[r][][0.6]{$q_4$}
\psfrag{Q5}[r][][0.6]{$q_5$}
\psfrag{C0}[r][][0.7]{$C_0$}
\psfrag{C1}[r][][0.7]{$C_1$}
\psfrag{C2}[r][][0.7]{$C_2$}
\psfrag{C3}[r][][0.7]{$C_3$}
\psfrag{C4}[r][][0.7]{$C_4$}
\psfrag{C5}[r][][0.7]{$C_5$}
\psfrag{C6}[r][][0.7]{$C_6$}
\psfrag{C7}[r][][0.7]{$C_7$}
\includegraphics[width=9cm]{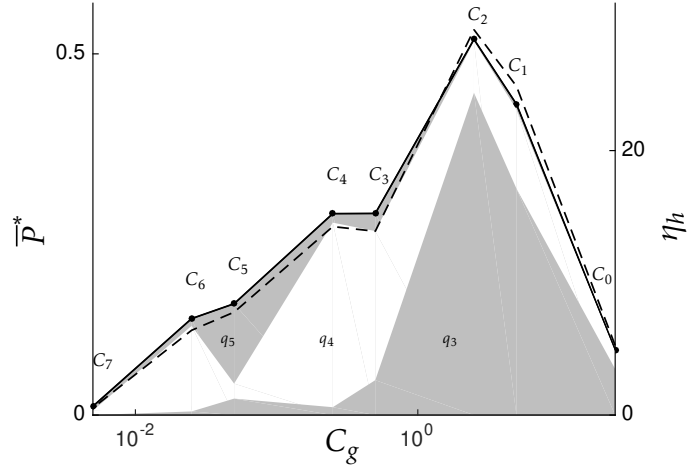}
\caption{Energy harvesting in vortex wakes: mean power coefficient (solid) and efficiency (dashed) versus damping coefficient; the distribution among the joints of the mean harvested power is denoted by the shaded areas.\label{fig:Powerharv}}
\end{figure}

Table \ref{tablePower} gathers the values of the mean power recovered by the dampers, the mean power spent to constrain the longitudinal position of the structures -- or equivalently to tow the cylinder-harvester in the ambient fluid -- and the harvesting efficiency as a function of the damping coefficient. Figure~\ref{fig:Powerharv} highlights the dependency between the mean recovered power and the damping coefficient of the system. We can observe that the maximum power is generated for the $C_2$ case where $C_g = 2.5$, corresponding to a harvesting efficiency of $29.14\%$. 

The distribution of the harvested power among the joints is also highlighted and the resulting values are presented in Table \ref{tableDistri}. The power distribution varies substantially depending on the case tested and the trend is non-trivial: the optimum configuration $C_2$ relies mostly on the first joint while the softer $C_4$ configuration exploits the second joint. This hints at a potential optimization which we leave as a topic of future work. % \sect{sec:optimhar}. 
Finally, if we consider that the cylinder cross-section is the representative scale of our device, we can see that the cylinder-harvester system can extract energy at a rate ($\overline{P}^*_{harvest, max}=0.52$) that is comparable to Betz' optimum for wind turbines ($\overline{P}^*_{Betz}=0.59$).

%\begin{tiny}
\begin{table}
\begin{center}
\resizebox{\textwidth}{!}{%
\begin{tabular}{c||c|c|c|c|c|c|c|c|c}
Case & $C_0$ & $C_1$ & $C_2$ & $C_3$ &  $C_4$ & $C_5$ & $C_6$ & $C_7$ & $C_8$\\
\hline $C_g$ & 25 &  5 &  2.5 &  0.5 &  0.25 &  0.05 &  0.025 &  0.005 & 0\\
\hline $\overline{P}^*_{harvest}$ & 0.090 & 0.430  & 0.520 & 0.279 & 0.279 & 0.154 & 0.134 & 0.012 & 0\\ 
\hline $\overline{P}^*_{F}$ & 1.71 & 1.73  & 1.78 & 2.01  & 1.96 & 1.98 & 2.08 & 2.25 & 0 \\  % remettre P_D-P_{harvest}
\hline $\eta_h$ & 5.26$\%$ & 24.87$\%$  & 29.14$\%$ & 13.90$\%$ &14.25$\%$ & 7.78$\%$ & 6.41$\%$ & 0.54$\%$ & 0\\
\end{tabular}}
\end{center}
\caption{Energy harvesting in vortex wakes: dimensionless mean power coefficients for several $C_g$ values.\label{tablePower}}
\end{table}
%\end{tiny}

%%\begin{tiny}
%\begin{table}
%\begin{center}
%\begin{tabular}{c||c|c|c|c|c}
%\multicolumn{2}{c|}{Case} & $C_0$ & $C_1$ & $C_2$ & $C_3$ \\
%\hline \multicolumn{2}{c|}{$C_g$} & 25 &  5 &  2.5 &  0.5 \\
%\hline \multicolumn{2}{c|}{$\overline{P}^*_{harvest}$} & 0.090 & 0.430  & 0.520 & 0.279 \\ 
%\hline \multicolumn{2}{c|}{$\overline{P}^*_{F}$} & 1.71 & 1.73  & 1.78 & 2.01 \\  % remettre P_D-P_{harvest}
%\hline \multicolumn{2}{c|}{$\eta_h$} & 5.26$\%$ & 24.87$\%$  & 29.14$\%$ & 13.90$\%$ \\
%\hline
%\hline Case &  $C_4$ & $C_5$ & $C_6$ & $C_7$ & $C_8$\\
%\hline  $C_g$ &  0.25 &  0.05 &  0.025 &  0.005 & 0 \\
%\hline  $\overline{P}^*_{harvest}$  & 0.279 & 0.154 & 0.134 & 0.012 & 0 \\ 
%\hline  $\overline{P}^*_{F}$ & 1.96 & 1.98 & 2.08 & 2.25 & 0 \\  % remettre P_D-P_{harvest}
%\hline  $\eta_h$ & 14.25$\%$ & 7.78$\%$ & 6.41$\%$ & 0.54$\%$ & 0 
%\end{tabular}
%\end{center}
%\caption{Energy harvesting in vortex wakes: dimensionless mean power coefficients for several $C_g$ values.\label{tablePower}}
%\end{table}
%%\end{tiny}

\begin{tiny}
\begin{table}
\begin{center}
\resizebox{\textwidth}{!}{%
\begin{tabular}{c||c|c|c|c|c|c|c|c|c}
Case & $C_0$ & $C_1$ & $C_2$ & $C_3$ &  $C_4$ & $C_5$ & $C_6$ & $C_7$ & $C_8$\\
\hline $C_g$ & 25 &  5 &  2.5 &  0.5 &  0.25 &  0.05 &  0.025 &  0.005 & 0\\
\hline $\overline{P}^*_{q_3}$ & 70.67 $\%$ & 72.85  $\%$ & 85.75 $\%$ & 17.32 $\%$ & 3.91 $\%$ & 14.78 $\%$ & 3.77 $\%$ & 0.62 $\%$ & 0 \\ 
\hline $\overline{P}^*_{q_4}$ & 27.64 $\%$ & 25.45  $\%$ & 13.69 $\%$ & 73.92 $\%$ & 91.62 $\%$ & 13.50 $\%$ & 88.72 $\%$ & 50.27 $\%$ & 0\\ 
\hline $\overline{P}^*_{q_5}$ & 1.68 $\%$ & 1.70  $\%$ & 0.56 $\%$ & 8.76 $\%$ & 4.48 $\%$ & 71.72 $\%$ & 7.51 $\%$ & 49.12 $\%$ & 0\\ 
\end{tabular}}
\end{center}
\caption{Energy harvesting in vortex wakes: distribution of the mean harvested power among the joints for all the $C_g$ values.\label{tableDistri}}
\end{table}
\end{tiny}

\subsubsection{Dynamics}
The resulting hydrodynamic forces can be studied from two perspectives, considering either (i) the harvester and the cylinder as two separate entities, and thus the effects of the harvester presence on the cylinder forces, or (ii) the cylinder-harvester system as a whole.

\begin{figure}[tbh] 
\center
\subfigure[Drag coefficient]{
			\psfrag{ylabel}[r][][0.8]{$C_{D,cyl/tot}$} 
			\psfrag{xlabel}[r][][0.8]{$t^*$}
			\includegraphics[width=5.5cm]{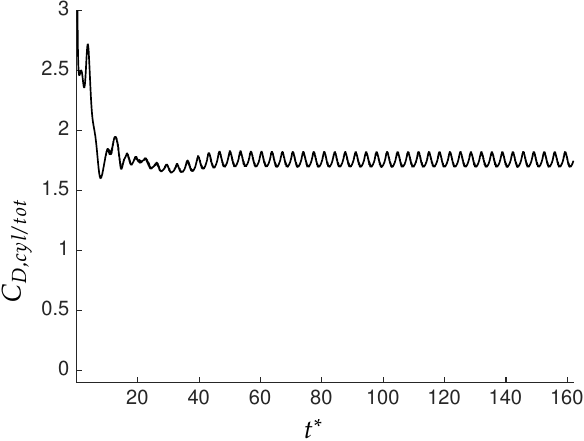}\label{fig:evola}}
\subfigure[Lift Coefficient]{
			\psfrag{ylabel}[r][][0.8]{$C_{L,cyl/tot}$} 
			\psfrag{xlabel}[c][][0.8]{$t^*$}
			\includegraphics[width=5.5cm]{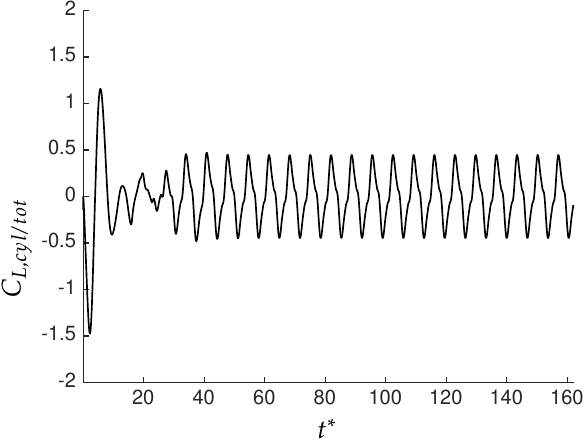}\label{fig:evolb}}
\caption{Energy harvesting in vortex wakes: histories of the drag and lift coefficient of the system for the $C_2$ case.\label{fig:evol}}
\end{figure}

\begin{figure}[tbh] 
\center
\subfigure[Mean drag coefficient]{
\psfrag{C0}[r][][0.7]{$C_0$}
\psfrag{C1}[r][][0.7]{$C_1$}
\psfrag{C2}[r][][0.7]{$C_2$}
\psfrag{C3}[r][][0.7]{$C_3$}
\psfrag{C4}[r][][0.7]{$C_4$}
\psfrag{C5}[r][][0.7]{$C_5$}
\psfrag{C6}[r][][0.7]{$C_6$}
\psfrag{C7}[r][][0.7]{$C_7$}
\psfrag{x}[c][][0.8]{$C_{g}$} 
\psfrag{y1}[c][][0.8]{$\overline{C}_{D,cyl}$}
\psfrag{y2}[c][][0.8]{$\Delta \overline{C}_{D,cyl} [\%]$}
\includegraphics[width=5.5cm]{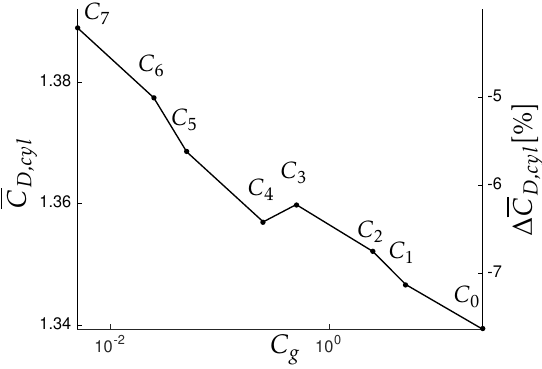}\label{fig:CD}}
\subfigure[Oscillatory part of the drag coefficient]{
\psfrag{x}[r][][0.8]{$C_{g}$} 
\psfrag{y1}[c][][0.8]{$C'_{D,cyl}$}
\psfrag{y2}[c][][0.8]{$\Delta C'_{D,cyl} [\%]$}
\includegraphics[width=5.5cm]{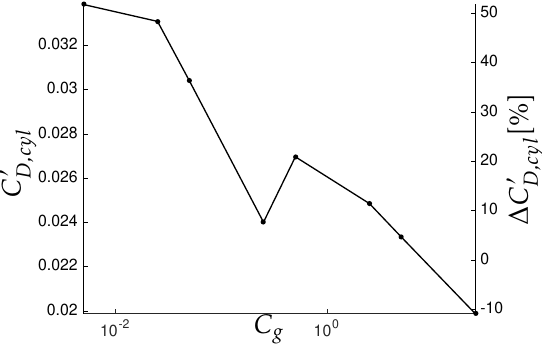}\label{fig:CD_a}}
\subfigure[Oscillatory part of the lift coefficient]{
\psfrag{x}[r][][0.8]{$C_{g}$} 
\psfrag{y1}[c][][0.8]{$C'_{L,cyl}$}
\psfrag{y2}[c][][0.8]{$\Delta C'_{L,cyl} [\%]$}
\includegraphics[width=5.5cm]{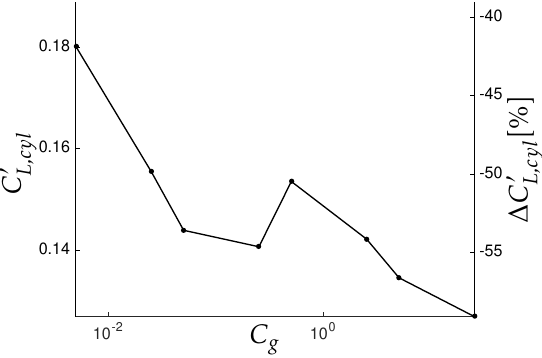}\label{fig:CL}}
\subfigure[Strouhal number of the cylinder (red) and harvester (black)]{
\psfrag{y}[r][][0.8]{$St$} 
\psfrag{x}[rt][][0.8]{$C_g$}
\includegraphics[width=5.1cm]{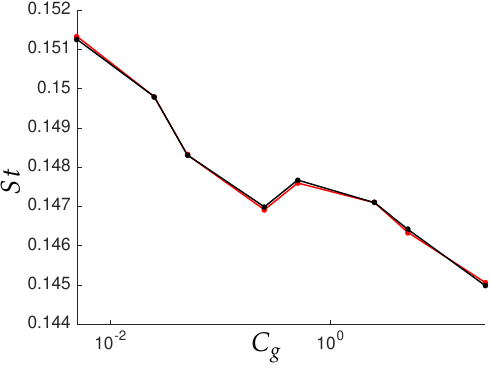}\label{fig:Strou}}
\caption{Energy harvesting in vortex wakes: evolutions of the cylinder force coefficients, also in terms of the relative differences with respect to an isolated cylinder, and of the Strouhal number versus the damping coefficient.\label{fig:CLCD}}
\end{figure}

\paragraph{Cylinder forces}
We consider the drag and lift coefficients, and more specifically, the amplitude of their oscillatory parts $C'_{L,\textrm{cyl}}\ \&\ C'_{D,\textrm{cyl}}$ and their means $\overline{C}_{L,cyl}\ \&\ \overline{C}_{D,\textrm{cyl}}$ (Table~\ref{tablecylinder}). Figure~\ref{fig:CLCD} shows the evolution of these coefficients according to the damping coefficient. The mean drag coefficient only decreases slightly whereas the oscillatory component appears to go through a more substantial reduction over the investigated damping coefficients. For the lift force, its mean remains at small values for all the configurations while its oscillatory part exhibits a notable reduction.
This very small mean lift and the periodicity of the oscillatory part (a representative history is provided in \fig{fig:evol}) indicate that the found asymptotic regimes are symmetric: there is no bias in the kinematics or dynamics in the transverse direction.

All these force parameters decrease with the damping coefficient but in a non-monotonic fashion, with a local-minimum for the configuration $C_4$. If we compare those values to those of an isolated cylinder, the \emph{no harvester} case of Table~\ref{tablecylinder}, we can observe decreases of up to $7.5\%$ for the mean drag coefficient and of $58\%$ for the oscillatory lift coefficient; however, the oscillatory part of the drag coefficient exhibits an increase up to $61\%$. 
\begin{table}[ht]
\begin{center}
\resizebox{\textwidth}{!}{%
\begin{tabular}{c||c|c|c|c|c|c|c|c|c|c}
Case & No harvester & $C_0$ & $C_1$ & $C_2$ & $C_3$ & $C_4$ & $C_5$ & $C_6$ & $C_7$ & $C_8$\\
\hline  $C_g$ & /  & 25 &  5 &  2.5 &  0.5 &  0.25 &  0.05 &  0.025 &  0.005 & 0 \\
\hline  $\overline{C}_{D,cyl}$ & 1.45 & 1.34 & 1.35 & 1.35 & 1.36 & 1.36 & 1.37 & 1.38 & 1.39 & 1.39 \\
\hline  $C'_{D,cyl} \times 10^2$ & 2.23 & 1.99 & 2.33 & 2.49 & 2.70 & 2.40 & 3.04 & 3.31 & 3.39 & 3.37\\
\hline  $\overline{C}_{L,cyl} \times 10^2$ & -0.54 & 0.13 & 0.11 & 0.08 & -0.02 & -0.03 & 0.09 & -0.01 & -0.31 & -0.43 \\
\hline  $C'_{L,cyl}$ & 0.31 & 0.13 & 0.13 & 0.14 & 0.15 & 0.14 & 0.14 & 0.16 & 0.18 & 0.19 \\
\end{tabular}}
\end{center}
\caption{Energy harvesting in vortex wakes: force coefficients of the cylinder for all the $C_g$ values.\label{tablecylinder}}
\end{table}

These modified hydrodynamic forces for the cylinder also translate in a modification of the shedding cycle of the cylinder. Figure~\ref{fig:Strou} shows the Strouhal numbers computed for the cylinder and the harvester, $S_{t,\textrm{cyl/har}} = \frac{f_{\textrm{cyl/har}}\,D}{U}$ where the frequency is extracted either from the lift coefficient of the cylinder $f_{\textrm{cyl}}$ or in the motion pattern of the harvester $f_{\textrm{har}}$. The two bodies get essentially synchronized; we can therefore consider a single Strouhal number, which covers the range $[0.145;0.151]$. These values are only slightly smaller than the one that was computed for the isolated cylinder $S_t = 0.154$.
%To summarize, this test case thus far shows that the cylinder-harvester system can extract energy at a rate ($\overline{P}^*_{harvest, max}=0.52$) that is comparable to Betz' optimum for wind turbines ($\overline{P}^*_{Betz}=0.59$); that extraction results into reduced force oscillations for the cylinder but does not modify its Strouhal number significantly.

\begin{figure}[tbh] 
\center
\subfigure[Mean drag coefficient]{
\psfrag{C0}[r][][0.7]{$C_0$}
\psfrag{C1}[r][][0.7]{$C_1$}
\psfrag{C2}[r][][0.7]{$C_2$}
\psfrag{C3}[r][][0.7]{$C_3$}
\psfrag{C4}[r][][0.7]{$C_4$}
\psfrag{C5}[r][][0.7]{$C_5$}
\psfrag{C6}[r][][0.7]{$C_6$}
\psfrag{C7}[r][][0.7]{$C_7$}
\psfrag{x}[c][][0.8]{$C_{g}$} 
\psfrag{y1}[c][][0.8]{$\overline{C}_{D,\textrm{sys}}$}
\psfrag{y2}[c][][0.8]{$\Delta \overline{C}_{D,\textrm{sys}} [\%]$}
\includegraphics[width=5.5cm]{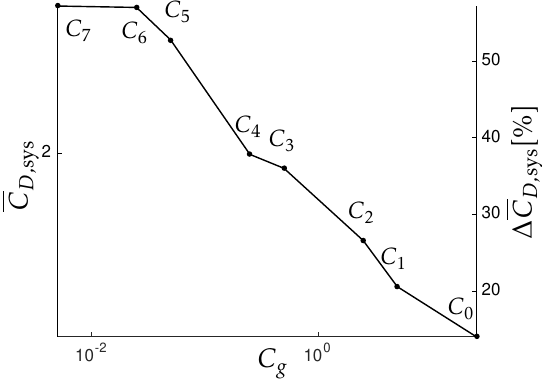}\label{fig:CDtot}}
\subfigure[Oscillatory part of the drag coefficient]{
\psfrag{x}[r][][0.8]{$C_{g}$} 
\psfrag{y1}[c][][0.8]{$C'_{D,\textrm{sys}}$}
\psfrag{y2}[c][][0.8]{$\Delta C'_{D,\textrm{sys}} [\%]$}
\includegraphics[width=5.5cm]{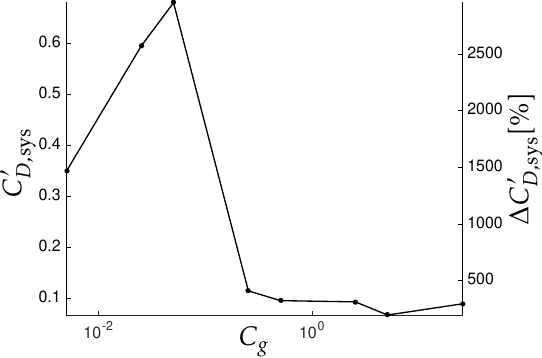}\label{fig:CDtot_a}}
\subfigure[Oscillatory part of the lift coefficient]{
\psfrag{x}[r][][0.8]{$C_{g}$} 
\psfrag{y1}[c][][0.8]{$C'_{L,\textrm{sys}}$}
\psfrag{y2}[c][][0.8]{$\Delta C'_{L,\textrm{sys}} [\%]$}
\includegraphics[width=5.5cm]{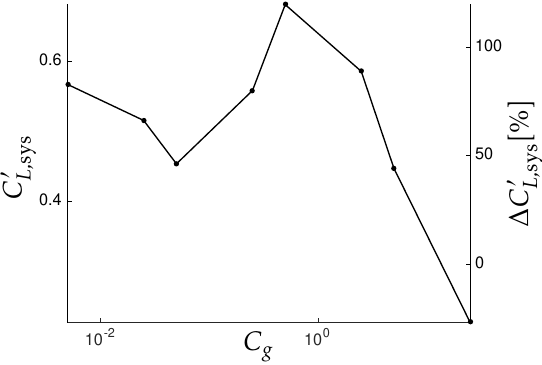}\label{fig:CLtot}}
\caption{Energy harvesting in vortex wakes: evolutions of the system force coefficients, also in terms of the relative differences with respect to an isolated cylinder, versus the damping coefficient.\label{fig:CLCDtot}}
\end{figure}

\paragraph{System forces}
Figure~\ref{fig:CLCDtot} and Table~\ref{tablesystem} give the evolution of the system force coefficients with damping. The mean total drag coefficient shows a clear decrease over the investigated range of damping coefficients; this drag is larger than for an isolated cylinder for all the cases and specifically, this difference amounts to $25\%$ for the optimal harvesting case ($C_2$ ). 
The oscillatory component appears to go through a more complex evolution. In particular, the oscillatory drag can show a $30$-fold increase for the lower damping values. For the lift force, its mean remains at small values for all the configurations while its oscillatory part exhibits a more intricate course. All cases increase the lift oscillations with respect to the isolated cylinder (up to $110\%$) with the exception of the case $C_0$ where the oscillatory part is reduced by $26\%$: the harvester is then quite stiff and tends to act as a fin.
\begin{table}[ht]
\begin{center}
\resizebox{\textwidth}{!}{%
\begin{tabular}{c||c|c|c|c|c|c|c|c|c|c}
Case & No harvester & $C_0$ & $C_1$ & $C_2$ & $C_3$ & $C_4$ & $C_5$ & $C_6$ & $C_7$ & $C_8$\\
\hline  $C_g$ & /  & 25 &  5 &  2.5 &  0.5 &  0.25 &  0.05 &  0.025 &  0.005 & 0 \\
\hline  $\overline{C}_{D,\textrm{sys}}$ & 1.45 & 1.65 & 1.75 & 1.84 & 1.97& 2.00 & 2.21 & 2.28 & 2.28 & 2.27  \\
\hline  $C'_{D,\textrm{sys}} \times 10^2$ & 2.23 & 8.86 & 6.65 & 9.25 & 9.48& 11.41 & 68.07 & 59.59 & 34.98 & 27.64\\
\hline  $\overline{C}_{L,\textrm{sys}} \times 10^2$ & -0.54 & 0.26 & 1.01 & -0.31 & -0.04& -0.74 & 0.34 & 0.45 & -1.07 & -1.59 \\
\hline  $C'_{L,\textrm{sys}}$ & 0.31 & 0.23 & 0.45 & 0.59 & 0.68 & 0.56 & 0.45 & 0.52 & 0.57 & 0.58 \\
\end{tabular}}
\end{center}
\caption{Energy harvesting in vortex wakes: force coefficients of the system for all the $C_g$ values.\label{tablesystem}}
\end{table}

 \section{Conclusions \label{sec:concl}}
We have developed a Fluid-Structure-Actuation Interaction solver for free-swimming problems based on the combination of a Vortex Particle-Mesh method with a Multi-Body System solver. 
Because it needs to time-integrate the dynamics of the actuated system too, the solver constitutes a marked departure from the force-free approach originally proposed in the works of~\cite{Coquerelle:2008,Gazzola:2011,Patankar:2005} and thus enables investigations of structural internal dynamics including complex kinematic constraints (e.g. rotational joints, linear rails, \textit{etc}). 
It indeed has to recover hydrodynamic forces and moments; this is achieved through a procedure that exploits the results of the intermediate steps of penalization and projection.
 
The resulting approach was applied to benchmark cases of increasing complexity: the sedimentation of a 2D cylinder, the flow past an elastically-mounted circular cylinder, a free swimming articulated fish and the problem of passive propulsion within the wake of a bluff body. The method is shown to reproduce reference results and accurately capture the coupled dynamics of the flow and the articulated structure.
The approach was then applied to the motion of an articulated eel-like harvester in the wake of a cylinder where the harvesting process is achieved by damper-like elements.
We performed a series of simulations to identify a maximum power recovery and concurrently studied the interaction between the fluid mechanics of the cylinder-harvester system and the power recovery. Our test configuration, in spite of its simple geometry, is an effective test-bed for the investigation of dynamically-rich harvesting problems. It has indeed shown some interesting features: a tuned uniform damping allows the harvester to achieve $88\%$ of Betz's optimum while a spatially-varying damping allows to harvest energy more uniformly over the device. These performances are promising as they hint at the efficient combination of power production and flow control, through drag reduction and lift stabilization. 

The present work focused on the Direct Numerical Simulation of moderate Reynolds number cases but nothing precludes the present methodology from higher Reynolds number cases. One should however use care in the handling of the penalization and projection techniques: (i) penalization will likely have to be quite stiff to remove spurious flows inside the body (one might then use an iterative penalization as in Hejlesen et al. \cite{Hejlesen:2015} and Gillis et al. \cite{Gillis:2017}); (ii) the grid will be finer in order to ensure that the thickness of the mollified region is small compared to the boundary layer around the body; (iii) the accuracy of the momentum flux, as calculated over the pseudo-time step of the projection, will probably lead to more severe constraints on the spatial and temporal resolutions. On the topic of time integration, we note that the method in its current state achieves a low temporal convergence order. One could either extend the method to multi-step schemes by using intermediate evaluations of the projection and penalization terms or use the above-mentioned iterative penalization to better capture added mass effects.

The present methodological contribution has a broader scope than vortex particle methods. Indeed, the hydrodynamic force extraction procedure can readily be transposed to the original velocity-pressure-based penalization context, as introduced by Angot et al.~\cite{Angot:1999}. It does also constitute a potential breakthrough for the robust handling of actuated and deformable structures, either position- or torque-controlled. It can be readily applied to the study of the interactions between a system, either artificial or biological, and its fluid environment. This allows to envision applications for the design of robotic swimmers and the control of their actuation, the study of neuromuscoluskeletal systems and gait generation, \textit{etc}.

%Our contributions provide a potential breakthrough in the sense that simplified fish-like structures can now be either position- or torque-controlled, while a dynamic model of biological or artificial actuators could be handily coupled. We claim that this is a critical step for designing bio-inspired efficient swimming agents, and to study closed-loop interactions between a fish and its environment. 
 
%Perspectives
Ongoing work focus on the extension to slender continuously elastic and actuated bodies. We note that the present approach makes the implementation step from two to three dimensions quite straightforward, as it does not entail any stencil modification or deep methodological change.
On the application side, this tool will now be leveraged in the investigation of control schemes and motion generation mechanisms to produce efficient and robust swimming gaits.

%
%%Perspectives
%Development work is ongoing for this tool. Just as for lift, one can account for drag through the shedding of the appropriate sheet-like structure. This will enable to account for stalled airfoils or blunt sections of blades. This should further improve our capture of the near wake as this drag signature is  at the center of helical instabilities inside shed vortices.
%Among other perspectives, we intend to leverage the high-quality LES results produced by this tool to design coarse rotor models for coarse LES of wind farms flows. The approach is evidently also well-suited for aeronautical applications such as aircraft or rotorcraft wakes. 
\section{Acknowledgements}
The present research benefited from computational resources made available on the Tier-1 supercomputer of the F\'ed\'eration Wallonie-Bruxelles, infrastructure funded by the Walloon Region under the grant agreement n$^o$1117545.
Caroline Bernier was funded by a FSR fellowship of Universit\'e catholique de Louvain for the project COMPACTSWIM (FSR 2013).\\
The authors thank the Blue Waters project (Grant Nos. OCI-0725070 and ACI-1238993), a joint effort of the University of Illinois at Urbana-Champaign and its National Center for Supercomputing Applications, for partial support (M.G.).
%!TEX root = JCP_bernier.tex
\appendix
\newpage
\section{Vortex method for fluid structure interaction  \label{app:former}}

In this Appendix, more details are given on the main computational step fo the vortex method formerly developed by \cite{Gazzola:2011}.

\subsection{Brinkman penalization details \label{app:brink}}
The immersed object shapes are described by a mollified characteristic function, $\chi_s$. This function is built upon a level set function that specifies the signed distance to the surface of the body, $d$. The mollified characteristic function is evaluated as
\begin{equation}
\label{eq:chi}
\chi_s = \left\lbrace  \begin{matrix} 0 & d<-\epsilon\\ 1/2[1+\frac{d}{\epsilon} + \frac{1}{\pi}\sin(\pi\frac{d}{\epsilon})] & |d| \leq \epsilon \\ 1 & d>\epsilon \end{matrix}\right. 
\end{equation}
where $\epsilon$ is the mollification length. For moderate Reynolds numbers, $\epsilon$ should be a small fraction ( about 1\%) of the characteristic length of the body geometry.
This geometry information is carried by a specific deformable grid; this allows to interpolate the color function onto the computational mesh in a flexible manner (see \cite{Gazzola:2011} for further details).

% MAttia thinks this is too long, should be cut shorter and maybe put a longer version or the details in appendix
\subsection{Splitting}
For the sake of completeness, all the operations performed in this method are summarized in the global Algorithm \ref{Algo1}. In this Algorithm, a first order Godunov splitting approach is used in order to solve \eq{eq:vorticity_pen}.  Given the penalized vorticity field \eq{eq:penalization}, the first step is to solve for the baroclinic term \eq{eq:baroclinic} then diffuse the vorticity strength \eq{eq:diffusion} and finally advect  the particules \eq{eq:advection}. It also highlights and explains the force-free character in the treatment of the fluid-structure interaction, as it handled as an exchange of momentum, or integrals of the fluid forces over a time step (\eqs{eq:proj1}-\eqref{eq:proj2}). 

%= \mathcal{P}_l (\bx , \mathcal{R}(\theta^n) \br_\mathcal{M}^n + \bx_{cm}^n,\chi_\mathcal{M}^n) and \mathcal{P}_l (\bx , \mathcal{R}(\theta^n) \br_\mathcal{M}^n + \bx_{cm}^n, \mathcal{R}(\theta^n) \dot{\br}_\mathcal{M}^n)
\mathleft
\begin{algorithm}
\caption{Original method\label{Algo1}}
\begin{scriptsize}
%\begin{algorithmic}
WHILE {$t^n \leq T_{end}$}
\begin{align}
 &\chi_s^n\text{ given deformation} \nonumber \\
 &\bu_{DEF}^n\text{ given deformation velocity field}\nonumber\\
 &\sigma = \chi_s^n(\nabla\cdot\bu_{DEF}^n)\nonumber\\
 &\rho^n = \rho_s^n\chi_s^n + \rho_f(1-\chi_s^n)\label{eq:density}\\
 &\nabla^2 \psi^n = -\omega^n\nonumber\\
 &\nabla^2 \phi^n = \sigma^n\nonumber\\
 &\bu^n = \nabla \times \psi^n+\nabla\phi^n\nonumber\\
 &\bu^n_T = \frac{1}{M_s}\int_\Sigma{\rho^n\chi_s^n\bu^n_*\ d\bx}\label{eq:proj1}\\
 &\dot{\theta}^n= \frac{1}{J^n_s}\int_\Sigma{\rho^n\chi_s^n(\bx-\bx_{cm}^n)\times\bu^n_*\ d\bx}\label{eq:proj2}\\
 &\bu_R^n = \dot{\theta}^n\times(\bx-\bx^n_{cm})\\
 &\bu_\lambda^n = \frac{\bu^n+\lambda\Delta t \chi_s^n(\bu^n_T+\bu^n_R+\bu^n_{DEF})}{1+\lambda\Delta t^n\chi_s^n}\\
 &\omega_\lambda^n = \nabla\times\bu_\lambda^n\label{eq:penalization}\\
 &\frac{\partial\omega^n_\lambda}{\partial t} = -\frac{\nabla\rho^n}{\rho^n}\times\left(\frac{\partial\bu^n_\lambda}{\partial t}+(\bu_\lambda^n\cdot\nabla)\bu_\lambda^n\right)\label{eq:baroclinic}\\
 &\frac{\partial\omega^n_\lambda}{\partial t} = \nu \nabla^2\omega_\lambda^n\label{eq:diffusion}\\
 &\frac{\partial\omega^n_\lambda}{\partial t} +\nabla	\cdot (\bu^n_\lambda\omega_\lambda^n	)=0\label{eq:advection}\\
 &\omega^{n+1} = \omega_\lambda^{n+1}\nonumber \\
 &\bx_{cm}^{n+1} = \bx_{cm}^n + \bu_T^n\Delta t^n\nonumber\\
 &\theta^{n+1} = \theta^n + \dot{\theta}^n\Delta t^n\nonumber\\
 &t^{n+1} = t^n+\Delta t^n\nonumber
\end{align}
ENDWHILE
\end{scriptsize}
%\end{algorithmic}
\end{algorithm}
\mathcenter

%!TEX root = JCP_bernier.tex
\newpage
\section{Hydrodynamic forces  \label{app:hydro}}

In this section, the hydrodynamic contact forces between the body and the fluid are computed. We first focus on the effort exerted by the fluid at the boundary of a volume $\Omega$. The resulting force and moment are expressed as

\begin{align}
\nonumber\bF^c &= \int_{\delta\Omega} \sigma \cdot \mathbf{n}dS;\\
\nonumber\bM^c &= \int_{\delta\Omega} \bx \times(\sigma \cdot \mathbf{n})dS.
\end{align}

Through the Green's theorem, both expression $\bF^c$ and $\bM^c$ are readily simplified as

\begin{align}
\nonumber\bF^c & = \int_{\delta\Omega} \sigma \cdot \mathbf{n}dS\\
 & = \int_{\Omega} \nabla \cdot \sigma dV;\label{eq:Fc}
\end{align}
and 
\begin{align}
\nonumber\bM^c & = \int_{\delta \Omega} \bx\ \times\ (\sigma \cdot {\bf n})dC \\
\nonumber M_i^c & = \int_{\delta \Omega}(\varepsilon_{ijk}\ x_j\ n_l\ \sigma_{lk})dC \\
\nonumber& \text{Through the Green's theorem, the expression becomes} \\
\nonumber& = \int_\Omega \frac{\partial}{\partial x_l} (\varepsilon_{ijk}\ x_j\ \sigma_{lk})dV\\
\nonumber& = \int_\Omega \varepsilon_{ijk} \frac{\partial x_j}{\partial x_l}\ \sigma_{lk}\ dV + \int_\Omega \varepsilon_{ijk}x_j \frac{\partial \sigma_{lk}}{\partial x_l}\ dV  \\
\nonumber& \text{The first term vanishes by symetry of the strain tensor, }\sigma\text{, i.e.} \\
\bM^c & = \int_{\Omega} \bx\ \times\ (\nabla \cdot \sigma)dV.\label{eq:Mc}
\end{align}

We now focus on the Navier-Stockes equations in their conservative form

\begin{equation}
\rho_f \frac{D \bu }{Dt} = \nabla \cdot \sigma	+ \rho_f \lambda (\bu_s-\bu). \label{eq:conserNS}
\end{equation}

The combination of \eqs{eq:Fc}, \eqref{eq:Mc} and \eqref{eq:conserNS} gives the following expressions:

\begin{align}
\nonumber\bF^c &=  \int_{\Omega} \left(\rho_f \frac{D \bu }{Dt} +  \rho_f \lambda (\bu-\bu_s)\right) dV\\
		& = \frac{d}{dt} \int_{\Omega_{mat}} \left(\rho_f \bu \right) dV+  \int_{\Omega} \left(\rho_f \lambda (\bu-\bu_s)\right) dV;\\
\nonumber\bM^c & = \int_{\Omega} \bx\ \times\ \left(\rho_f \frac{D \bu }{Dt} +  \rho_f \lambda (\bu-\bu_s)\right)dV \\
\nonumber	&  \text{Following the Reynolds transport theorem, it becomes } \\
		& =  \frac{d}{dt} \int_{\Omega_{mat}} \bx\ \times\ \left(\rho_f \bu \right)dV+  \int_{\Omega} \bx\ \times\ \left(\rho_f \lambda (\bu-\bu_s)\right)dV
\end{align}

And we retrieve the equations used in Section \ref{sec:method}: \eqs{eq:Fhyd} and \eqref{eq:Mhyd}.

%\section*{References}

%\bibliography{Bibliographie}

\end{document}